\documentclass{pasj00}
\draft

\begin{document}
\SetRunningHead{S.Bessho and T.Tsuribe}{Fragmentation of Filament under UV}
%\Received{}%{yyyy/mm/dd}
%\Accepted{}%{yyyy/mm/dd}
%\Published{}%{yyyy/mm/dd}

\title{Fragmentation of Primordial Filamentary Clouds under Far-Ultraviolet Radiation}

%%% begin:list of authors
% Do NOT capitalize all letters in "textsc".
\author{Shinji \textsc{Bessho} and Toru \textsc{Tsuribe}} %
 %\thanks{Example: Present Address is xxxxxxxxxx}}
\affil{Department of Earth and Space Science, Osaka University, Machikaneyama 1-1, Toyonaka, Osaka 560-0043}
\email{bessho@vega.ess.sci.osaka-u.ac.jp}
%%% end:list of authors

%%% Please use the following style in case that sorting by 
%%% affilation is impossible. 
%
% \author{%
%   D-Firstname \textsc{D-Familyname}\altaffilmark{1}
%   E-Firstname \textsc{E-Familyname}\altaffilmark{1,2}
%   and
%   F-Firstname \textsc{F-Familyname}\altaffilmark{2}}
% \altaffiltext{1}{Address of Institute}
% \email{ddddd@xxx.xxx.xx.xx}
% \email{eeeee@xxx.xxx.xx.xx}
% \altaffiltext{2}{Address of Institute}

%% `\KeyWords{}' always has to be placed before `\maketitle'.
 %Do NOT move this preamble from here!

\maketitle

\begin{abstract}
Collapse and fragmentation of uniform filamentary clouds under isotropic far-ultraviolet external radiation are investigated. Especially, impact of photodissociation of hydrogen molecules during collapse is investigated. Dynamical and thermal evolution of collapsing filamentary clouds are calculated by solving virial equation and energy equation with taking into accounts non-equilibrium chemical reactions. It is found that thermal evolution is hardly affected by the external radiation if the initial density is high ($n_0 > 10^2 \mathrm{cm^{-3}}$). On the other hand, if line mass of the filamentary cloud is moderate and initial density is low ($n_0 \le 10^2 \mathrm{cm^{-3}}$), thermal evolution of the filamentary cloud tends to be adiabatic owing to the effect of the external dissociation radiation. In this case, collapse of the filamentary cloud is suppressed and the filamentary cloud fragments into very massive clouds ($\sim 10^{4-5} M_\odot$) in the early stage of collapse. Analytic criterion for the filamentary clouds to fragment into such massive clouds is discussed. We also investigate collapse and fragmentation of the filamentary clouds with an improved model. This model can partly capture the effect of run-away collapse. Also in this model, the filamentary clouds with low initial density ($n_0 \le 10^2 \mathrm{cm^{-3}}$) fragment into massive clouds ($\sim 10^4 M_\odot$) owing to the effect of the external radiation.
\end{abstract}

\section{Introduction}
It is widely accepted that in the dark age the first collapsing density perturbation collapses and cools owing to hydrogen molecules ($\mathrm{H_2}$) to form so-called population III (popIII) stars (Bromm et al. 1999, 2002; Abel et al. 2000, 2002; Yoshida et al. 2008). PopIII is expected to form in halos typically with $\sim 10^6 M_\odot$ (Tegmark et al. 1997). If popIIIs are massive stars, they are expected to cause feedbacks via supernovae and radiation. The former may spread metals made by nuclear fusion and sweeps neighboring gas by shock (e.g., Heger et al. 2003). This paper focuses on the latter. There are two types of the radiative feedbacks, which are ionization and dissociation (Whalen et al. 2004; Susa et al. 2009; Hasegawa et al. 2009a, 2009b). Ionization of hydrogen atoms ($\mathrm{H}$) provides strong heating which causes evaporation of clouds. However, since inter stellar matter mainly consists of $\mathrm{H}$ with large opacity, ionization photon tends to be prevented from spreading, and photoionization would occur mainly within halos. On the other hand, photodissociation of $\mathrm{H_2}$ would occur even out of halos (Kitayama et al. 2004). Thus, it is expected that there are some regions which are not photoionized but photodissociated. We investigate gravitational collapse and fragmentation of primordial clouds in such a region.

Since $\mathrm{H_2}$ is main coolant in the early universe, if $\mathrm{H_2}$ is photodissociated, a collapsing primordial cloud ($> 10^8 M_\odot$) heats adiabatically up to high temperature ($\sim 10^4 \mathrm{K}$) where atomic cooling becomes effective. In such a case, it is suggested that high mass objects ($\sim 10^{4-5} M_\odot$) may form because Jeans mass ($M_J \propto T^{3/2} \rho^{-1/2}$) is large (e.g., Bromm $\&$ Loeb 2003). This possibility may connect the direct formation of supermassive black hole. However, if collapsing cloud fragments into many small clumps, collapsed object may form star cluster (Omukai $\&$ Yoshii 2003). In order to understand actual final outcome and their initial mass function clearly, further detailed investigation of thermal and dynamical evolution of collapsing clouds is required.

Omukai (2001) investigated the evolution of spherical clouds under the dissociation radiation. The author calculated the evolution of the central region of a collapsing sphere, assuming free-fall collapse. Started from $n_0=8.9 \times 10^{-2} \mathrm{cm}^{-1}$, the clouds collapse adiabatically in the early stage, $n \le 10^2 \mathrm{cm^{-3}}$. After this stage, thermal evolution of clouds is divided into two types of tracks. When the external radiation is very strong (e.g, with the intensity larger than $10^{-18} \mathrm{erg\ cm^{-2} s^{-1} Hz^{-1} sr^{-1}}$ at $13.6\mathrm{eV}$ for thermal radiation of $10^4 \mathrm{K}$), $\mathrm{H_2}$ is photodissociated enough to suppress $\mathrm{H_2}$ cooling. In this case, main coolant is provided by hydrogen atom. On the other hand, if the intensity of the external radiation is moderate, sufficient amount of $\mathrm{H_2}$ forms and clouds cool mainly via $\mathrm{H_2}$ cooling. Susa (2007) investigated more realistic evolution of a spherical cloud under the UV radiation from a single light source by three-dimensional calculations. The author investigated whether or not clouds collapse for parameters such as distance from the light source and the density when the light source turns on.

As for the formation of the spherical clouds investigated above, a filamentary cloud is a possible origin. Filamentary clouds are commonly expected during the way to form the stars. When sheet-like cloud forms, the sheet-like cloud tends to fragment into the filamentary clouds (Miyama, Narita, \& Hayashi 1987a, b). In cosmological simulation, the filamentary structure forms from density perturbation which has $\ge 10^6 M_\odot$ (Abel et al. 1998; Bromm et al. 1999; Greif et al. 2008). These filamentary clouds have a possibility to produce spherical clouds by fragmentation (Nagasawa 1987; Inutsuka $\&$ Miyama 1997). Thus, in order to understand the origin and property of a collapsing spherical cloud, it is useful to investigate the evolution of filamentary clouds. Previous works about fragmentation of primordial filamentary clouds include one-zone models (Uehara et al. 1996; Flower 2002; Omukai \& Yoshii 2003) and one-dimensional models (Nakamura \& Umemura 1999, 2001, 2002; Uehara $\&$ Inutsuka 2000). Among these studies, Uehara et al. (1996), Omukai \& Yoshii (2003), and Nakamura \& Umemura (1999, 2001) considered only $\mathrm{H_2}$ as a coolant, and the others considered deuterated hydrogen molecules ($\mathrm{HD}$) as well as $\mathrm{H_2}$. If the initial fraction of $\mathrm{H_2}$ is lower than $10^{-3}$, $\mathrm{H_2}$ becomes main coolant and the fragment mass is $1 - 500 M_\odot$. On the other hand, if the initial fraction of $\mathrm{H_2}$ is higher than $3 \times 10^{-3}$, a filamentary cloud with low initial density ($n_0 < 10^4 \mathrm{cm^{-3}}$) cools mainly via $\mathrm{HD}$ cooling since $\mathrm{H_2}$ promotes the formation of $\mathrm{HD}$ with larger cooling rate. In this case, the fragment mass is $1 - 140 M_\odot$. When $\mathrm{HD}$ becomes main coolant, the filamentary clouds cool down to lower temperature (e.g., $\sim 40 \mathrm{K}$) and fragment into less massive fragments\footnote{Since low temperature helps collapse, the timescale of density evolution is short and fragmentation does not occur until density becomes very high ($\sim 10^{12} \mathrm{cm^{-3}}$) (see $\S 2.3$).} than when $\mathrm{H_2}$ is main coolant. All of these previous papers except Omukai $\&$ Yoshii (2003) did not consider the feedback effect from the external radiation.

Once a massive star forms in a cloud $\sim 10^6 M_\odot$, the whole cloud is photodissociated by UV radiation from the star (Omukai $\&$ Nishi 1999). In order to form subsequent stars, larger cloud e.g., $\ge 10^8 M_\odot$ is required. Omukai \& Yoshii (2003) investigated fragmentation of the filamentary cloud under the UV radiation using a one-zone model. Based on the results by Nakamura \& Umemura (2001), they assumed that the filamentary cloud fragments when its density becomes 100 times higher than that at the loitering point at which temperature is a local minimum in $\rho - T$ plane owing to $\mathrm{H_2}$ cooling. The authors concluded that the fragment mass is smaller under the stronger UV radiation. However, the condition for fragmentation given by Nakamura \& Umemura (2001) is considered for the cases without the external radiation. Thus, it is not clear that this condition for fragmentation is applicable for the cases with the external radiation. When the filamentary cloud suffers photodissociation, it will collapse adiabatically and it will fragment before it reaches loitering point. In such a case, the fragment mass becomes larger than in the cases without the external radiation.

In this paper, we investigate the effect of photodissociation radiation on the thermal evolution of the collapsing filamentary cloud and on the mass of fragment. We investigate whether or not the external radiation photodissociates enough $\mathrm{H_2}$ to prevent $\mathrm{H_2}$ cooling and whether or not the external radiation makes fragment mass larger. Instead of assuming free-fall collapse, we calculate the dynamical evolution by solving virial equation taking into account the effect of pressure gradient force as a result of insufficient cooling due to photodissociation. As for the condition for fragmentation, we assume that the filamentary clouds fragment when the timescale of fragmentation becomes shorter than the timescale of density evolution (see $\S 2.3$).

We measure the effect of the external radiation with respect to fragment mass. We investigate the dependence of fragment mass on parameters such as initial density, line mass, and intensity of the external radiation. We also discuss analytically what physical process determines fragment mass. We consider the simplest case in which the dissociation radiation is isotropic. As for the intensity of the external radiation originating from forming popIIIs, we refer the result by Dijkstra et al. (2008) which investigated the mean intensity of the external dissociation radiation in the universe at $z \sim 10$.

In $\S 2$, we describe the model which is used in this paper. We present numerical and analytic results in $\S 3$ and $\S 4$, respectively. We present an improved model and numerical results in $\S 5$. Finally, $\S 6$ is devoted to summary and discussion, including discussion about evolution of cloud after fragmentation.

\section{Model}
\subsection{Basic equations}
For simplicity, we assume that the filamentary clouds are uniform. We solve the virial equation for the dynamical evolution in the cylindrical radial direction (Uehara et al. 1996). We do not consider dark matter for simplicity. In the case with high initial density, baryon density is expected to dominate dark matter density and it will give a good approximation. In the case with the low initial density, we will underestimate the effect of dark matter, e.g., large infall velocity owing to dark matter gravity. The virial equation for the filamentary cloud of unit length with volume $V$ is
\begin{eqnarray}
\frac{1}{2} \frac{d^2 I}{dt^2}=2 \Psi +2\Pi -Gl^2 ,
\end{eqnarray}
where $G$ is gravitational constant, $l$ is the line mass (mass per unit length) of the filamentary cloud,
\begin{eqnarray}
I=\int_V \rho r^2 dV
\end{eqnarray}
is the inertial moment per unit length with density $\rho$ and radius $r$,
\begin{eqnarray}
\Psi=\int_V \frac{1}{2}\rho v^2 dV
\end{eqnarray}
is the kinetic energy per unit length with the velocity $v$, and
\begin{eqnarray}
\Pi=\int_V P dV
\end{eqnarray}
is the integrated pressure per unit length with local pressure $P$. By substituting equations (2)-(4) into equation (1), we have
\begin{eqnarray}
\frac{d^2 R}{dt^2}=-\frac{2G}{R} \{ l-l_c(T) \} ,
\end{eqnarray}
where $R$ is the radius of the filamentary cloud, $T$ is temperature, and $l_c(T)$ is the critical line mass for the hydro-static isothermal filamentary cloud defined as
\begin{eqnarray}
l_c(T) \equiv \frac{2k_B T}{\mu m_{\mathrm{H}}G}
\end{eqnarray}
(Ostriker 1964) with Boltzmann constant $k_B$, mean molecular weight $\mu$, and mass of a hydrogen atom $m_{\mathrm{H}}$.

For the thermal evolution, we solve energy equation
\begin{eqnarray}
\frac{du}{dt}= -P\frac{d\ }{dt} \frac{1}{\rho}-\frac{\Lambda_{\mathrm{radi}}}{\rho}-\frac{\Lambda_{\mathrm{chem}}}{\rho},
\end{eqnarray}
where $u$ is thermal energy per unit mass
\begin{eqnarray}
u=\frac{1}{\gamma_{\mathrm{ad}}-1} \frac{k_B T}{\mu m_{\mathrm{H}}}
\end{eqnarray}
with adiabatic exponent $\gamma_{\mathrm{ad}}$. The first term of the right hand side in equation (7) denotes adiabatic heating. The radiative cooling rate $\Lambda_{\mathrm{radi}}$ per unit volume includes lines of $\mathrm{H}$, lines of $\mathrm{H_2}$, lines of $\mathrm{HD}$, and continuum (see table1 in detail; see also Omukai 2001). For the radiative cooling, the effect of radiative transfer is included according to Susa et al. (1996). The symbol $\Lambda_{\mathrm{chem}}$ represents heating/cooling rate associated with chemical reactions. Equation of state for ideal gas
\begin{eqnarray}
P=\frac{\rho k_B T}{\mu m_{\mathrm{H}}}
\end{eqnarray}
is assumed.

We consider non-equilibrium chemical reactions by solving equations
\begin{eqnarray}
\frac{df_i}{dt}=\sum_{j,\ k} k_{ijk} f_j f_k n + \sum_j k_{ij} f_j ,
\end{eqnarray}
where $n$ is number density of all nuclei, $k_{ijk}$ and $k_{ij}$ are reaction rates of formation and destruction of species $i$, and $f_i$ is the fraction of species $i$. We consider the following fourteen species : $\mathrm{H}$, $\mathrm{H^+}$, $\mathrm{H^-}$, $\mathrm{H_2}$, $\mathrm{H_2^+}$, $\mathrm{He}$, $\mathrm{He^+}$, $\mathrm{He^{++}}$, $\mathrm{D}$, $\mathrm{D^+}$, $\mathrm{D^-}$, $\mathrm{HD}$, $\mathrm{HD^+}$, and $\mathrm{e^-}$. We consider 35 reactions concerned with $\mathrm{H}$ and $\mathrm{He}$ taken from Omukai (2001) and 18 reactions concerned with $\mathrm{H}$ and $\mathrm{D}$ taken from Nakamura $\&$ Umemura (2002). We also consider photodestruction of species $\mathrm{D}$, $\mathrm{D^-}$, and $\mathrm{HD^+}$ (we refer to Galli $\&$ Palla 1998 for $\mathrm{D}$ and $\mathrm{HD^+}$ and Frolov 2004 for $\mathrm{D^-}$). Above equations are solved numerically with implicit integrator.

\subsection{External radiation}
We assume the external radiation to be isotropic. Dijkstra et al. (2008) investigated the mean intensity of the dissociation radiation at $z \sim 10$ from the surrounding star-forming halos and estimated the probability distribution of the mean intensity. We adopt mean intensities whose probabilities are $\sim 0.4$ \footnote{$0.4$ is the highest probability.} and $\sim 0.06$ in Dijkstra et al. (2008) (see $\S 2.4$). Moreover we assume that the external radiation is thermal radiation from $120M_\odot$ stars and surface temperature ($T_{\mathrm{sur}}=95719\mathrm{K}$) of stars is determined according to Schaerer (2002). We also assume that the ionization photon does not reach the filamentary clouds from light sources.

When we calculate photodissociation reaction
\begin{eqnarray}
\mathrm{H_2} + \gamma \rightarrow \mathrm{H_2}^\ast \rightarrow 2 \mathrm{H}
\end{eqnarray}
(solomon process)\footnote{$\mathrm{H_2}^\ast$ is excited $\mathrm{H_2}$. $\gamma$ is photon with $12.4\mathrm{eV}$.}, we consider the extinction of photon by dissociation of $\mathrm{H_2}$ and absorption by continuum processes (see Table.1). The photodissociation rate is proportional to the mean intensity of the dissociation radiation. During penetrating the filamentary cloud from the surface to the center, the intensity of the dissociation radiation decreases owing to dissociation of $\mathrm{H_2}$ and absorption by continuum processes. In this paper, the effects of radiative transfer of dissociation photon is approximated by the product of shield factor as $J_\nu=f_{\mathrm{sh}} f_{\mathrm{con}} J_{\nu,0}$, where $f_{\mathrm{sh}}$ is self-shielding factor associated with photodissociation of $\mathrm{H_2}$, $f_{\mathrm{con}}$ is decreasing rate associated with absorption by continuum processes, and $J_{\nu,0}$ is the mean intensity of the dissociation radiation at the surface of the filamentary cloud.

First, we consider the photon decreasing rate, $f_{\mathrm{con}}$, associated with the absorption of dissociation photon by continuum processes. We focus on the dissociation photons with $12.4\mathrm{eV}$. Radiative transfer equation along the $s$-direction is given by
\begin{eqnarray}
\frac{dI_\nu}{ds}=-k_\nu I_\nu + j_\nu ,
\end{eqnarray}
where $I_\nu$ is the intensity of the radiation of frequency $\nu$, $k_\nu$ is the total opacity associated with reactions in Table.1 (see Appendix 1), and $j_\nu$ is emissivity. For simplicity we assume that the scattered photons are absorbed immediately. We also assume $j_\nu = 0$ since $j_\nu$ mainly consists of lines and continuum with lower energy than $12.4\mathrm{eV}$. We consider the length of the column in various directions. Using the length $R/\sin \theta$ from surface to the center of the filamentary cloud with angle $\theta$ from the axis of the filamentary cloud, the intensity of the external radiation at the center is given by
\begin{eqnarray}
I_\nu(0)=J_{\nu,0} \exp \biggl( -\frac{k_\nu R}{\sin \theta} \biggr).
\end{eqnarray}
Hence, $f_{\mathrm{con}} J_{\nu,0}$ is given by
\begin{eqnarray}
f_{\mathrm{con}} J_{\nu,0}=\frac{J_{\nu,0}}{4\pi} \int^{2 \pi}_{0} d \phi \int^{\pi}_{0} d \theta \ \sin \theta \exp \biggl( -\frac{k_\nu R}{\sin \theta} \biggr).
\end{eqnarray}
The integral of the right hand side in equation (14) is calculated with the fitted functions given in Appendix 2. In typical results shown in $\S 3$, the value of $f_{\mathrm{con}}$ is found to be larger than $0.97$ during collapse for $n \le 10^2 \mathrm{cm^{-3}}$, where the photodissociation of $\mathrm{H_2}$ is effective. The absorption in the low density cloud has only a minor effect.

Second, we consider self-shielding factor, $f_{\mathrm{sh}}$, associated with the dissociation. The self-shielding factor $f_{\mathrm{sh}}$ is approximated by
\begin{eqnarray}
f_{\mathrm{sh}}=\mathrm{min} \biggl[ 1,\ \biggl(\frac{N_{\mathrm{H_2}}}{10^{14} \mathrm{cm^{-2}}} \biggr)^{-3/4} \biggr]
\end{eqnarray}
(Draine $\&$ Bertoldi 1996), where $N_{\mathrm{H_2}}$ is the column density of $\mathrm{H_2}$. To estimate the effective column density, we estimate average in angle of the length between the surface and the center. Using the length $R/\sin \theta$ from the surface to the center of the filamentary cloud in the direction with angle $\theta$ from the axis of the filamentary cloud, effective column density of the filamentary cloud is estimated as
\begin{eqnarray}
N_{\mathrm{H_2}} = \frac{1}{4\pi} \int^{2 \pi}_{0} d \phi \int^{\pi}_{0} d \theta \  \sin \theta \frac{R}{\sin \theta} n_{\mathrm{H_2}} = \frac{\pi}{2} n_{\mathrm{H_2}}R ,
\end{eqnarray}
where $n_{\mathrm{H_2}}$ is number density of $\mathrm{H_2}$. Since $R \propto n^{-1/2}$, effective column density increases during collapse as $N_{\mathrm{H_2}} \propto n_{\mathrm{H_2}}R \propto n^{1/2}$. Finally, the photodissociation reaction rate of $\mathrm{H_2}$ is given by
\begin{eqnarray}
k_{\mathrm{2step}}=1.4 \times 10^9 f_{\mathrm{con}} f_{\mathrm{sh}} J_{\nu,0} \ \mathrm{s^{-1}}.
\end{eqnarray}

\subsection{Fragmentation of filamentary clouds}
During collapse of the filamentary cloud, two important timescales exist. One is the timescale of density evolution defined as $t_{\mathrm{dyn}} \equiv \rho_c/\dot{\rho_c}$ where $\rho_c$ is the density at the center. The other is the timescale of fragmentation defined as $t_{\mathrm{frag}} \equiv 2.1 / \sqrt{2 \pi G \rho_c}$ (Nagasawa 1987; Uehara et al. 1996). The latter is the timescale in which the fastest growing mode of perturbation grows to non-linear. According to Inutsuka $\&$ Miyama (1992), if acceleration in the radial direction is high, perturbation with low amplitude does not grow enough during collapse. When acceleration in the radial direction becomes low owing to strong pressure, $t_{\mathrm{dyn}}$ becomes large and the growth of perturbation becomes remarkable. The wave length of the fastest growing mode of perturbation is nearly the diameter of the filamentary clouds (Nagasawa 1987). This wave length becomes shorter during collapse. If the fastest growing mode has time to grow to non-linear before the diameter of the filamentary cloud changes largely, the filamentary cloud is expected to fragment. Thus, we assume that the filamentary clouds start to fragment at the moment when $t_{\mathrm{frag}}<t_{\mathrm{dyn}}$ is satisfied (Uehara et al. 1996; Inutsuka $\&$ Miyama 1997).

Using the wave length of the fastest growing mode $\lambda_{\mathrm{frag}} \sim 2 \pi R$ at fragmentation, the fragment mass is estimated as
\begin{eqnarray}
M_{\mathrm{frag}} \equiv \lambda_{\mathrm{frag}} l \sim 2 \pi R l
\end{eqnarray}
(Narita, Miyama, $\&$ Hayashi 1987a, b; Larson 1985; Uehara et al. 1996). According to equation (18), the fragment mass is proportional to the radius of the filamentary clouds. If fragmentation occurs after the filamentary cloud collapses to a small radius with high density, the fragment mass is small.

\subsection{Parameters and initial conditions}
In this paper, we treat three physical quantities as parameters, which are initial number density $n_0$, normalized intensity of the external radiation,
\begin{eqnarray}
J_{21} \equiv \frac{J_{h\nu=13.6\mathrm{eV},0}}{10^{-21} \mathrm{erg\ cm^{-2} s^{-1} Hz^{-1} sr^{-1}}},
\end{eqnarray}
and the line mass parameter,
\begin{eqnarray}
f \equiv \frac{\pi G \rho_0 \mu m_{\mathrm{H}}}{2k_B T_0} R_0^2 = \frac{l}{l_c(T_0)}
\end{eqnarray}
with initial density $\rho_0$, initial temperature $T_0$, and initial radius $R_0$ \footnote{The typical value of $f$ is $2$. This value is realized when the sheet-like gas fragments by the fastest growth rate (Miyama et al. 1987a).}. The reason why we choose these three quantities is as follows: in the view point of dynamical evolution, the line mass parameter $f$ is important. In the view point of thermal evolution, initial density $n_0$ is important. The symbol $J_{\nu,0}$ and $n_0$ are necessary to study the effect of dissociation photon.

We consider cases with $\log_{10} n_0=1$, $1.5$, $2$, $2.5$, $3$, $3.5$, $4$, $4.5$, $5$, $5.5$, and $6$ for $n_0$ and $f=1.25$, $1.5$, $1.75$, $2$, $2.25$, $2.5$, $2.75$, and $3$ for $f$. For $J_{21}$ we consider $J_{21}=1$, $6.5$, and $10$. The case with $J_{21}=1$ demonstrates the weak external radiation. According to Dijkstra et al.(2008), $J_{21}=6.5$ is the average intensity at $z \sim 10$, and $J_{21}=10$ represents strong radiation case whose probability is $0.06$ (see $\S 2.2$).

We assume that radial infall velocity at the surface of the filamentary cloud equals to the sound speed. If the filamentary clouds form from the sheet-like cloud, gravitational force dominates pressure gradient force in the filamentary cloud. Thus, the infall velocity when the filamentary cloud forms is expected to be in the same order as the sound speed, i.e., $v = \alpha c_s$ with a numerical coefficient $\alpha \sim o(1)$ which depends on the details of fragmentation. As a typical value, we set $\alpha=1$ according to Nakamura $\&$ Umemura (2002). We calculated several cases with various values of $\alpha$ and found that the evolution hardly changes for $\alpha < 5$. We mention the case with $\alpha=5$ in $\S 3.4$.

In this paper, the filamentary cloud is assumed to form from a cloud which experiences $\mathrm{H_2}$ cooling without UV radiation. We also assume that the external radiation turns on when the filamentary clouds form. The initial values of temperature and $f_{\mathrm{H_2}}$ are set to be $300 \mathrm{K}$ and $f_{\mathrm{H_2}}= 10^{-4}$. In addition to $f$, $n_0$, and $J_{21}$, for the thermal evolution, $T_0$ and fraction of $\mathrm{H_2}$, $f_{\mathrm{H_2}}$, are also important. We adopt one typical value for them. This value of $f_{\mathrm{H_2}}$ is typically seen in cosmological simulations (e.g., Abel et al. 1998), $f_\mathrm{H_2} \sim 10^{-4}-10^{-3}$. We discuss initial $\mathrm{H_2}$ fraction with the effect of the dissociation radiation in $\S 3.5$. Fraction of $\mathrm{He}$, $f_{\mathrm{He}}$, is set to be $0.0825$ which corresponds to the mass fraction $Y_p=0.244$ (Izotov $\&$ Thuan 1998). Initial fraction of electron, $f_e$, is set to be $10^{-4}$ according to Uehara et al. (1996). We adopt this value of $f_e$ in order for electron not to change $f_{\mathrm{H_2}}$ largely in the early stage of collapse\footnote{Electron helps $\mathrm{H_2}$ formation via $\mathrm{H^-}$ channel.}. Initial fraction of proton is determined from the charge conservation. We assume that $[\mathrm{D}]/[\mathrm{H}]=4 \times 10^{-5}$, which is consistent with observations of the deuterium Ly$\alpha$ feature in the absorption spectra of high-redshift quasars (e.g., O'Meara et al. 2001). Fraction of the other species is set to be zero at the initial state.

%%%%%%%%%%%%%%%%%%%
\section{Results}
%%%%%%%%%%%%%%%%%%%
%------------------------------------------------%
\subsection{Cases without the external radiation}
%------------------------------------------------%
\subsubsection{Low density filamentary clouds with large line mass}
%------------------------------------------------%

Figure 1 shows the result of the case with low initial density and large line mass, $(f,n_0,J_{21})=(3,10\mathrm{cm^{-3}},0)$. From the early stage of collapse, adiabatic heating rate is a little higher than $\mathrm{H_2}$ cooling rate and temperature gradually increases. At $n \sim 10^8 \mathrm{cm^{-3}}$, $l_c$ overcomes $l$, and the filamentary cloud begins to be decelerated. Above $n \sim 10^8 \mathrm{cm^{-3}}$, the three-body reaction becomes efficient and $\mathrm{H_2}$ fraction increases to $\sim 0.4$. At $n \sim 10^{10} \mathrm{cm^{-3}}$, although $\mathrm{H_2}$ cooling is still effective, chemical heating associated with the three-body reaction of $\mathrm{H_2}$ formation also becomes effective. Thus, temperature continues to increase. When the density reaches $n \sim 10^{11} \mathrm{cm^{-3}}$, although the three-body reaction is inefficient, temperature stops to increase owing to sufficient cooling with a large fraction of $\mathrm{H_2}$. The filamentary cloud becomes optically thick to $\mathrm{H_2}$ line emissions at $n \sim 10^{12} \mathrm{cm^{-3}}$. Around this density, temperature increases again and it eventually exceeds $2000\mathrm{K}$. Such a high temperature state causes collisional dissociation of $\mathrm{H_2}$. Since chemical cooling associated with this dissociation can not dominate adiabatic heating, temperature is kept high enough to decelerate collapse. As a result, the filamentary cloud fragments when density reaches $n \sim 10^{15} \mathrm{cm^{-3}}$. Several authors pointed out that $\mathrm{H_2}$ collision-induced emission becomes effective at $n \sim 10^{15} \mathrm{cm^{-3}}$ (Omukai $\&$ Nishi 1998; Ripamonti $\&$ Abel 2004; Yoshida et al. 2006). However, in the case in figure 1, since temperature is high ($\sim 3000 \mathrm{K}$), 80$\%$ of $\mathrm{H_2}$ is dissociated and cooling rate of $\mathrm{H_2}$ collision-induced emission is smaller than adiabatic heating rate by two orders of magnitude. Since the density of the filamentary cloud at fragmentation is very high ($n \sim 10^{15} \mathrm{cm^{-3}}$), the mass of fragment is small ($\sim 0.1 M_\odot$).

In summary, the evolution of the low density models with large line mass is affected largely by radiative cooling and chemical heating/cooling associated with $\mathrm{H_2}$. In this sense, our result is qualitatively same as the previous results by Uehara et al. (1996) and Nakamura $\&$ Umemura (1999, 2001, 2002). Note that the above result with sub-solar mass of fragment originates from the one-zone model with a uniform filamentary cloud. In $\S 5$, we show the result with an improved model with the effect of run-away collapse.

\subsubsection{High density filamentary clouds with small line mass}
Figure 2 shows the result of the case with high initial density and small line mass, $(f,n_0,J_{21})=(1.25,10^6\mathrm{cm^{-3}},0)$. In this case, adiabatic heating dominates cooling a little after the early stage of the collapse, $n \le 3 \times 10^6 \mathrm{cm^{-3}}$. Collapse is accelerated only in the early stage of collapse ($n \le 2 \times 10^6 \mathrm{cm^{-3}}$) and not after that. Since acceleration is limited in the short density range, collapse of the filamentary cloud is limited at lower density ($n \sim 10^8 \mathrm{cm^{-3}}$) and the fragment mass is larger ($\sim 50 M_\odot$) than the case in figure 1 ($\S 3.1.1$). Different from the case in figure 1, $\mathrm{H_2}$ cooling never dominates.

Figure 3 shows the fragment mass for various $n_0$ and $f$. All lines are similar to each other and can be approximated as $M_{\mathrm{frag}} \sim 230 n_0^{-0.03} f^{-5.1}$ with an error at most factor $4$ at $f=3$. This approximate function agrees with numerical results at low $f$ ($<2$). The fragment mass is determined mainly by $f$. This tendency agrees with the result of Uehara et al. (1996). Nakamura $\&$ Umemura (2002) suggested that the fragment mass depends mainly on $n_0$. However, our results do not agree with that of Nakamura $\&$ Umemura (2002). This difference comes from simplicity that the filamentary clouds is assumed to be uniform. In the uniform model, virial temperature is determined by the whole line mass ($f$($=l/l_c$))\footnote{Since collapse of the uniform filamentary cloud is homologous, virial temperature is determined by whole line mass, that is $f$. On the other hand, since collapse of the filamentary cloud is run-away collapse in one-dimensional model, virial temperature is determined by mass of the central region. The mass of the central region mainly depends on $n_0$.}. The evolution of the non-uniform filamentary cloud (e.g., the one-dimensional model) includes run-away characteristics of the flow. The improved model with the effect of run-away collapse will be introduced in $\S 5$.

%==============================================%
\subsection{Cases with the external radiation}
%==============================================%
%------------------------------------------------------------------%
\subsubsection{Low density filamentary clouds with large line mass}
%------------------------------------------------------------------%
Figure 4 shows the result of the case with low initial density, large line mass, and strong external radiation, $(f,n_0,J_{21})=(3,10\mathrm{cm^{-3}},10)$. The case in figure 4 corresponds to the case in figure 1 ($\S 3.1.1$) with the external radiation. In figure 4, it is seen that $f_{\mathrm{H_2}}$ decreases owing to photodissociation in the early stage of collapse, and that adiabatic heating dominates from the early stage of collapse. Cylindrical collapse is decelerated at $n \sim 10^2 \mathrm{cm^{-3}}$ since temperature increases. However, this deceleration is temporary and the filamentary cloud does not fragment at this point. Instead, it continues to collapse and shields itself from the dissociation photon. Then $f_{\mathrm{H_2}}$ begins to increase at $n \sim 10^2 \mathrm{cm^{-3}}$. After that, $\mathrm{H_2}$ cooling becomes efficient and the evolution becomes similar to that in figure 1 ($\S 3.1.1$). As a result, the filamentary cloud collapses until it becomes optically thick to $\mathrm{H_2}$ lines, and it fragments into the low mass clumps about $0.14 M_\odot$. This mass of fragments is expected to be underestimated owing to the uniform filament model with homologous collapse as in the case in figure 1 (\S 3.1.1).

%-------------------------------------------------------------------%
\subsubsection{Low density filamentary clouds with small line mass}
%-------------------------------------------------------------------%
Figure 5 shows the result of the case with low initial density, small line mass, and strong external radiation, $(f,n_0,J_{21})=(1.25,10\mathrm{cm^{-3}},10)$. In this case, the external radiation photodissociates $\mathrm{H_2}$ in the early phase since dissociation photon penetrates the filamentary cloud with low column density. The early photodissociation suppresses $\mathrm{H_2}$ cooling. As a result, temperature increases adiabatically until fragmentation. Since collapse is terminated and fragmentation occurs at low density ($\sim 34 \mathrm{cm^{-3}}$), the fragment mass is very large ($\sim 10^5 M_\odot$). The difference between cases with and without the external radiation is whether or not $\mathrm{H_2}$ is dissociated by the external radiation in the early stage. If $\mathrm{H_2}$ is photodissociated sufficiently enough to suppress $\mathrm{H_2}$ cooling, the filamentary clouds with low $f$ ($< 2.5$) evolve adiabatically and fragment into very massive clumps.

The difference between figure 5 and figure 4 ($\S 3.2.1$) is the value of line mass. Since the initial density is low, in both cases $\mathrm{H_2}$ is dissociated and the filamentary clouds evolve adiabatically in their initial stage. Since the line mass for the case in figure 5 is smaller, the slight increase of temperature is sufficient to suppress collapse. On the other hand, since the line mass for the case in figure 4 is larger, the slight increase of temperature is not sufficient to suppress collapse. Hence in the case in figure 4 the filamentary cloud does not fragment in the early stage of collapse and eventually $\mathrm{H_2}$ forms enough to cool the filamentary cloud. Difference of these two results originates from the line mass of the filamentary clouds. A critical line mass to shield themselves from the dissociation photon is discussed analytically in $\S 4.4$.

%-------------------------------------------------------------------%
\subsubsection{High density filamentary clouds with small line mass}
%-------------------------------------------------------------------%
Figure 6 shows the result of the case with high initial density, small line mass, and strong external radiation, $(f,n_0,J_{21})=(1.25,10^6\mathrm{cm^{-3}},10)$. The case in figure 6 corresponds to the case in figure 2 ($\S 3.1.2$) with the external radiation. The evolution of the filamentary cloud in figure 6 is similar to that in figure 2 ($\S 3.1.2$). This is because the initial density is high enough to shield the filamentary cloud from dissociation photon. Adiabatic heating dominates $\mathrm{H_2}$ cooling from the early stage of collapse, and temperature increases gradually. In this case, the filamentary cloud fragments into the slightly more massive fragments than in the case without the external radiation (figure 2 ($\S 3.1.2$)). This is because the external radiation dissociates a little $\mathrm{H_2}$ in the early stage of collapse. However, the difference is negligible.

%-------------------------------------------------------------------%
\subsubsection{Parameter dependence of temperature evolution}
%-------------------------------------------------------------------%
To investigate how parameters affect the evolution of temperature, we systematically calculate with changing one of three parameters in the parameter space ($n_0$, $f$, $J_{21}$). For unchanged parameters, $f=1.25$, $n_0=10 \mathrm{cm^{-3}}$, and $J_{21}=10$ are used.

Figure 7 shows the evolution of temperature in the cases with various $n_0$. It is seen that in the cases with $n_0=10-10^2 \mathrm{cm^{-3}}$ temperature increases adiabatically since $\mathrm{H_2}$ is photodissociated. However, with $n_0$ higher than $10^2 \mathrm{cm^{-3}}$, the filamentary clouds shield themselves from dissociation photon and cool owing to $\mathrm{H_2}$ cooling.

Figure 8 shows the evolution of temperature in the cases with various $f$. In all the cases, most of $\mathrm{H_2}$ is photodissociated in the early phase ($n < 10^2 \mathrm{cm^{-3}}$) and temperature increases adiabatically. Each line in figure 8 overlaps each other during the early stage of collapse. This is because the effect of photodissociation is similar for the same density. The fragment mass depends on $f$. The filamentary clouds with $f < 2.5$ fragment during the early adiabatic evolution ($n \le 10^2 \mathrm{cm^{-3}}$). In the case with large $f$, the filamentary clouds collapse to high density since virial temperature is large. The filamentary clouds with $f \ge 2.5$ form sufficient amount of $\mathrm{H_2}$ to cool during collapse even with strong dissociation radiation ($J_{21}=10$). Once the filamentary clouds cool, they continue to collapse and reach the high density ($n \ge 10^{13} \mathrm{cm^{-3}}$) before fragmentation.

Figure 9 shows the evolution of temperature in the cases with various $J_{21}$. In the cases with the external radiation ($J_{21} \ge 1$), temperature increases adiabatically since most of $\mathrm{H_2}$ is photodissociated and $\mathrm{H_2}$ cooling is suppressed. On the other hand, in the case without the external radiation ($J_{21}=0$), temperature does not increase adiabatically.

%=========================%
\subsection{Fragment mass}
%=========================%
Figure 10 shows the fragment mass for all the parameters by using contours maps in $n_0 - f$ plane. Results for different values of $J_{21}=0$, $1$, $6.5$, and $10$ are presented in different diagrams. Solid lines in each diagram of figure 10 represent constant fragment mass. The dotted line and the dash-dotted line will be referred in $\S 4$. In diagram b) with $J_{21}=1$, the region with large fragment mass ($> 10^4M_\odot$) is found in the range $n_0 \le 10^{1.5} \mathrm{cm^{-3}}$ and $f<1.5$. This region is clearly as the result of the external radiation since such a region does not exist in diagram a) with $J_{21}=0$. With larger $J_{21}$ in diagrams c) and d), it is seen that the region with massive fragment ($> 10^4M_\odot$) becomes larger in $n_0-f$ plane. In diagram c), this region spreads up to $n_0 \sim 10^2 \mathrm{cm^{-3}}$ and $f=2.25$. In diagram d), this region spreads up to $f=2.5$.

However, in the case with $n_0>10^2\mathrm{cm^{-3}}$ or $f \ge 2.5$, it is seen that the fragment mass is hardly changed by the external radiation. This is explained as follows : as for the cases with large $n_0$, the filamentary clouds shield themselves from the dissociation radiation from the early stage of evolution. As for the cases with large $f$, as shown in $\S 3.2.1$, the filamentary clouds continue to collapse up to density high enough to shield themselves from the external radiation and form $\mathrm{H_2}$ even if $\mathrm{H_2}$ is photodissociated in the early stage of collapse. The filamentary cloud in the uniform model whose collapse is homologous tends to collapse to higher density than the realistic model whose collapse is run-away collapse. Thus, it is probably as the result of our choice of the uniform model that the filamentary cloud with a little larger $f$ than moderate value collapses to high density in spite of $\mathrm{H_2}$ loss in the early stage of collapse. In $\S 5$, we compare the result with the modified one-zone model including the effect of run-away collapse. Further investigation including spatial variation will be presented in the separate paper.

%================================================%
\subsection{Effect of supersonic initial velocity}
%================================================%
We assumed initial infall velocity at the cloud surface to be same as the sound velocity. Here, we comment on the effect of faster initial infall velocity which may be possible under the effect of dark matter gravity. Much faster initial velocity can help the filamentary clouds to shield themselves from the dissociation radiation due to the rapid evolution of density before fragmentation. We have checked this possibility with the model as in figure 5 ($\S 3.2.2$) where most of $\mathrm{H_2}$ is photodissociated during the early stage of collapse. It is found that the filamentary cloud can collapse to form $\mathrm{H_2}$ and shield themselves from the dissociation radiation if the initial velocity is five times or lager than five times sound speed. An example of numerical result is shown in figure 11 where photodissociation is suppressed at the early stage of collapse and the filamentary cloud does not fragment until density becomes large ($\sim 10^8 \mathrm{cm^{-3}}$). Thus, the filamentary clouds with highly supersonic initial infall velocity tend to avoid the effect of the external radiation. This is consistent qualitatively with Hasegawa et al. (2009b) who investigated the formation of globular clusters as a result of fragmentation in such a way.

%================================================%
\subsection{Effect of photodissociation on the initial fraction of $\mathrm{H_2}$}
%================================================%
Although initial $\mathrm{H_2}$ fraction is assumed to be $10^{-4}$ at $n = n_0 \ge 10\mathrm{cm^{-3}}$ in \S 3 (and $\S 5$), this value of $\mathrm{H_2}$ fraction is expected to be affected by the external radiation before the density of the cloud reaches $n_0$. In previous papers without the external radiation, $f_{\mathrm{H_2}}=10^{-4}-10^{-3}$ is adopted as initial states after virialization. In this sense, the above initial value of $f_{\mathrm{H_2}}=10^{-4}$ should be regarded as for the case where the external radiation turns on at the moment when the filamentary cloud with $n=n_0$ forms. On the other hand, Omukai $\&$ Yoshii (2003) considered the cases with sufficiently low initial density $n_0=0.1\mathrm{cm^{-3}}$, in which the external radiation had turned on before the filamentary cloud forms.

Is our assumption that initial $f_{\mathrm{H_2}}$ is set to be $10^{-4}$ valid when the external radiation turns on before the filamentary cloud forms? In this subsection, we investigate how much $f_{\mathrm{H_2}}$ is at $n=10\mathrm{cm^{-3}}$ in the case where the external radiation turns on at $n_{\mathrm{UV}}$ which is lower than $10\mathrm{cm^{-3}}$. We calculate the evolution of $\mathrm{H_2}$ fraction by using the cloud with sufficiently low initial density $n_0=0.1\mathrm{cm^{-3}}$. Initial fraction of $\mathrm{H_2}$ is set to be zero. In figure 12, $\mathrm{H_2}$ fraction at $n=10\mathrm{cm^{-3}}$ is shown as a function of $n_{\mathrm{UV}}$ for the cases with different $J_{21}$. It is seen that $\mathrm{H_2}$ fraction at $n=10\mathrm{cm^{-3}}$ is much different between in the cases with different $J_{21}$ and $n_{\mathrm{UV}}$. To set $f_{\mathrm{H_2}}=10^{-4}$ at $n=n_0$ in $\S 3$ (and $\S 5$) is valid only when the external radiation turns on at the moment when the filamentary cloud forms.

%%%%%%%%%%%%%%%%%%%%%%%%%%%%%%%%%
\section{Analytic investigation}
%%%%%%%%%%%%%%%%%%%%%%%%%%%%%%%%%
In this section, we analytically investigate the property shown in the numerical results in $\S 3$. To explain the property of the collapsing filamentary cloud, three criteria are considered in the view point whether or not the filamentary cloud can cool during collapse.

%=====================================================================%
\subsection{Cooling criterion 1 : Whether cooling is effective or not}
%=====================================================================%
There is a critical value $n_a$ of initial density that determines whether or not $\mathrm{H_2}$ cooling dominates adiabatic heating at the start of collapse. Before we consider the effect of the external radiation, we derive $n_a$ without the external radiation. If initial density exceeds $n_a$, temperature increases from the early stage of collapse and the filamentary clouds fragment into massive fragments ($\sim 50 M_\odot$) as shown in figure 2 ($\S 3.1.2$). If $\mathrm{H_2}$ cooling dominates adiabatic heating at the start of collapse, the following inequality is satisfied :
\begin{eqnarray}
-P \frac{d\ }{dt}\frac{1}{\rho} < \frac{\Lambda_{\mathrm{H_2}}}{\rho}.
\end{eqnarray}
The rate of $\mathrm{H_2}$ cooling is approximated as
\begin{eqnarray}
\Lambda_{\mathrm{H_2}} \simeq \left\{ \begin{array}{ll}
2.5 \times 10^{-26} n^2 f_{\mathrm{H_2}} \displaystyle 
\biggl(\frac{T}{300 \mathrm{K}} \biggr)^3 & n \ll 10^4 \mathrm{cm^{-3}} \\
8.0 \times 10^{-24} n f_{\mathrm{H_2}} \displaystyle 
\biggl(\frac{T}{300 \mathrm{K}} \biggr)^{3.8} & n \gg 10^4 \mathrm{cm^{-3}} \\
\end{array} \right.
\end{eqnarray}
(Galli $\&$ Palla 1998), where $\Lambda_{\mathrm{H_2}}$ is in units of $\mathrm{erg\ cm^{-3} s^{-1}}$. Assuming that the timescale of collapse is the free-fall time ($1/\rho \cdot d\rho /dt = -t_{\mathrm{ff}}^{-1}$), equation (21) becomes
\begin{eqnarray}
\frac{k_B T_0}{\mu m_{\mathrm{H}}} \cdot \sqrt{2\pi G \rho} < \frac{\Lambda_{\mathrm{H_2}}}{\rho}
\end{eqnarray}
(see Appendix 3). Using $\Lambda_{\mathrm{H_2}}$ for $n \ll 10^4 \mathrm{cm^{-3}}$ in equation (22) with $f_{\mathrm{H_2}} = 10^{-4}$ for the initial state, the condition for cooling is found to be
\begin{eqnarray}
n > n_a \equiv 1.9 \times 10^2 \mathrm{cm^{-3}} 
\biggl( \frac{T_0}{300\mathrm{K}} \biggr)^{-6} 
\biggl( \frac{f_{\mathrm{H_2}}}{10^{-4}} \biggr)^{-2}.
\end{eqnarray}
On the other hand, in the case with $n \gg 10^4 \mathrm{cm^{-3}}$, adiabatic heating always dominates $\mathrm{H_2}$ cooling. Thus, $\mathrm{H_2}$ cooling dominates adiabatic heating at the start of collapse for $n_a < n_0 < 10^4 \mathrm{cm^{-3}}$. In figure 2, since the filamentary cloud has higher initial density than $10^4 \mathrm{cm^{-4}}$, temperature increases at the early stage of the collapse. Since this condition does not include the effect of the external radiation, equation (24) should be accepted as a necessary condition for cooling.

%=================================================%
\subsection{Equilibrium fraction of $\mathrm{H_2}$}
%=================================================%
Let us prepare to investigate the condition whether $\mathrm{H_2}$ cooling dominates adiabatic heating in the early stage with the effect of the external radiation. Since cooling rate depends on $\mathrm{H_2}$ fraction, we first estimate the equilibrium fraction of $\mathrm{H_2}$ which is attained when formation and photodissociation of $\mathrm{H_2}$ balance under the external radiation. Assuming the chemical equilibrium between formation and photodissociation of $\mathrm{H_2}$, the fraction of $\mathrm{H_2}$ is found to be
\begin{eqnarray}
f_{\mathrm{H_2}}=\frac{n f_e k_{\mathrm{H^-}}}{k_{\mathrm{2step}}} ,
\end{eqnarray}
where $k_{\mathrm{H^-}}=1.0 \times 10^{-18} T \ \mathrm{cm^3 s^{-1}}$ is the reaction rate for $\mathrm{H^-}$ channel,
\begin{eqnarray}
\mathrm{H}+\mathrm{H^-} \rightarrow \mathrm{H_2} + \mathrm{e^-}.
\end{eqnarray}
At the initial state in our model, timescale of formation of $\mathrm{H_2}$, $t_{\mathrm{form}}$, is given by
\begin{eqnarray}
t_{\mathrm{form}}=
\frac{1}{k_{\mathrm{H^-}} n_0 f_e} = 
3.33 \times 10^{18} \mathrm{s} 
\biggl(\frac{T}{300\mathrm{K}}\biggr)^{-1} 
\biggl(\frac{n_0}{10\mathrm{cm^{-3}}}\biggr)^{-1} 
\biggl(\frac{f_e}{10^{-4}}\biggr)^{-1}.
\end{eqnarray}
On the other hand, assuming that $N_{\mathrm{H_2}}$ is larger than $10^{14} \mathrm{cm^{-2}}$, timescale of photodissociation is given by
\begin{eqnarray}
t_{\mathrm{diss}}
=\frac{1}{k_{\mathrm{2step}} f_{\mathrm{H_2}}} = 2.26 \times 10^{17} \mathrm{s} 
\biggl(\frac{J_{21}}{1}\biggr)^{-1} 
\biggl(\frac{N_{\mathrm{H_2}}}{10^{14}\mathrm{cm^{-2}}}\biggr)^{3/4} 
\biggl(\frac{f_{\mathrm{H_2}}}{10^{-4}}\biggr)^{-1}.
\end{eqnarray}

Equilibrium $\mathrm{H_2}$ fraction, $f_{\mathrm{H_2,eq}}$, can be estimated by the condition $t_{\mathrm{form}} = t_{\mathrm{diss}}$. Substituting the column density
\begin{eqnarray}
N_{\mathrm{H_2}}&=&
\frac{\pi}{2} n f_{\mathrm{H_2}} R = 
\frac{\pi n^{1/2} f^{1/2} f_{\mathrm{H_2}}}{2 m_{\mathrm{H}}} \sqrt{ \frac{2 k_B T}{\pi \mu G}}\ \nonumber \\
&\sim& 2.40 \times 10^{21} \mathrm{cm^{-2}} 
\biggl( \frac{n}{10 \mathrm{cm^{-3}}} \biggr)^{1/2} 
\biggl( \frac{T}{300 \mathrm{K}} \biggr)^{1/2} f^{1/2} f_{\mathrm{H_2}}
\end{eqnarray}
into equation (28)\footnote{There are cases where $f_{\mathrm{H_2}} \sim 10^{-8}$ and $N_{\mathrm{H_2}} \sim 2.40 \times 10^{13} \mathrm{cm^{-2}} < 10^{14} \mathrm{cm^{-2}}$. However, in such cases, although we use $f_{\mathrm{sh}} = (N_{\mathrm{H_2}}/10^{14} \mathrm{cm^{-2}})^{-3/4}$, we do not face significant error.}, we have
\begin{eqnarray}
f_{\mathrm{H_2,eq}}=
\mathrm{min} \biggl[ 2.88 \times 10^{-5} 
\biggl(\frac{n}{10 \mathrm{cm^{-3}}} \biggr)^{11/2} 
\biggl(\frac{T}{300 \mathrm{K}} \biggr)^{11/2} J_{21}^{-4} f^{3/2},\ 1\ \biggr].
\end{eqnarray}
According to equation (30), $f_{\mathrm{H_2,eq}}$ is expected to be large for high $n$. In the cases with $J_{21}=1$, $6.5$, and $10$, equation (30) predicts $f_{\mathrm{H_2,eq}} \sim 1$ for $n > 76.6 \mathrm{cm^{-3}}$, $299\mathrm{cm^{-3}}$, and $409 \mathrm{cm^{-3}}$, respectively. Especially, $f_{\mathrm{H_2,eq}}$ is expected to be large enough in $n \gg 10^4 \mathrm{cm^{-3}}$ where $\mathrm{H_2}$ is hardly dissociated by the external radiation. In order to check the applicability of equation (30) to analytic criteria for massive fragment formation, we compared $f_{\mathrm{H_2}}$ by equation (30) with the numerical results of the evolution of the filamentary cloud. In the cases with $J_{21} = 6.5$ and $10$, it is found that $f_{\mathrm{H_2,eq}}$ agrees with numerical results within error of 40 $\%$. On the other hand, in the case with $J_{21} = 1$ where $\mathrm{H_2}$ formation dominates photodissociation, it is found that $f_{\mathrm{H_2}}$ given by equation (30) is about 2.5 orders of magnitude smaller than the numerical result.

%==============================================================================================================%
\subsection{
Cooling criterion 2 : 
Whether cooling becomes effective when formation and photodissociation of $\mathrm{H_2}$ balance
}
%==============================================================================================================%
In this subsection, we derive the condition whether $\mathrm{H_2}$ cooling dominates adiabatic heating in the early stage under the external radiation by assuming that the formation of $\mathrm{H_2}$ balances with photodissociation. Cooling time is estimated as
\begin{eqnarray}
t_{\mathrm{cool}}=\frac{3 n k_B T}{2 \Lambda_{\mathrm{H_2}}}.
\end{eqnarray}
On the other hand, free-fall time of the uniform filamentary cloud is given by 
\begin{eqnarray}
t_{\mathrm{ff}}=\frac{1}{\sqrt{2 \pi G \rho}}.
\end{eqnarray}
By equating $t_{\mathrm{cool}}$ and $t_{\mathrm{ff}}$ for $n \ll 10^4 \mathrm{cm^{-3}}$ with assuming with $f_{\mathrm{H_2}} < 1$, we have critical initial density $n_b$ as
\begin{eqnarray}
%f_c=31.4 \biggl(\frac{n}{10 \mathrm{cm^{-3}}} \biggr)^{-4} 
%         \biggl(\frac{T}{300 \mathrm{K}} \biggr)^{-17/3} J_{21}^{8/3}.
%n_b = 23.7 f^{-1/4} 
%          \biggl(\frac{T}{300 \mathrm{K}} \biggr)^{-17/12} 
%          J_{21}^{2/3}.
n_b = 78 \mathrm{cm^{-3}}
         \biggl(\frac{T}{300 \mathrm{K}} \biggr)^{-5/4} 
         \biggl(\frac{J_{21}}{10} \biggr)^{2/3}
         f^{-1/4}.
\end{eqnarray}
In the case with $n_0 > n_b$, $\mathrm{H_2}$ photodissociation is too weak to halt $\mathrm{H_2}$ cooling. On the other hand, the case with $n_0 < n_b$ has a possibility to halt $\mathrm{H_2}$ cooling. In diagrams b), c), and d) of figure 10, prediction by equation (33) is plotted by the dash-dotted lines\footnote{The dash-dotted line is not drown in diagram a) of figure 10 since we are interested only in the case with the external radiation.}. Comparing with numerical results, it is found that the dash-dotted line in figure 10 gives us a reasonable criterion for above two thermal evolutions.

%==============================================================================================================%
\subsection{Cooling criterion 3 : Whether cooling is effective with temperature adiabatically increasing}
%==============================================================================================================%
We investigate another criterion by which $\mathrm{H_2}$ cooling becomes effective during the collapse under the condition where temperature increases adiabatically as a result of strong photodissociation of $\mathrm{H_2}$ in the early stage. There are numerical examples presented in figure 4 ($\S 3.2.1$)/figure 5 ($\S 3.2.2$) where $\mathrm{H_2}$ cooling is effective/ineffective. Since in both examples most of $\mathrm{H_2}$ is once photodissociated, the difference between these two examples seems to be originated from the difference in line mass. Here, we derive the critical line mass $f_c$ for effective $\mathrm{H_2}$ cooling after strong photodissociation from the condition by which the filamentary cloud continues to collapse up to the density high enough to shield themselves from the dissociation photon.

We assume $n = n_b \ll 10^4 \mathrm{cm^{-3}}$ (i.e., formation and photodissociation of $\mathrm{H_2}$ balance). We define that $\mathrm{H_2}$ cooling is ``effective'' if $\mathrm{H_2}$ cooling dominates adiabatic heating when the right hand side of equation (5) equals zero (i.e., gravitational force balances with pressure gradient force). In the adiabatic evolution, temperature at the density $n$ is represented as
\begin{eqnarray}
T=T_0 \biggl(\frac{n}{n_0} \biggr)^{2/3}.
\end{eqnarray}
When the right hand side of equation (5) equals zero, we have
\begin{eqnarray}
\frac{2 k_B T}{\mu m_{\mathrm{H}} G}=l ( = f l_c(T_0) ).
\end{eqnarray}
Since the filamentary clouds mainly consist of hydrogen atom, we assume $\mu \sim 1$. Using equations (34) and (35), we have
\begin{eqnarray}
f=\frac{T}{T_0}.
\end{eqnarray}
Using equations (33), (34), and (36), we have
\begin{eqnarray}
f_c=
2.0 \biggl( \frac{n_0}{10 \mathrm{cm^{-3}}} \biggr)^{-1/3} 
    \biggl( \frac{T_0}{300 \mathrm{K}} \biggr)^{-5/12}
    \biggl(\frac{J_{21}}{10} \biggr)^{2/9}.
\end{eqnarray}
For the cases with $f < f_c$, $\mathrm{H_2}$ cooling never dominates adiabatic heating and the filamentary clouds fragment into very massive fragments ($\sim 10^{4-5} M_\odot$ ; c.f., figure 5 ($\S 3.2.2$)). In figure 10, the condition $f=f_c$ is shown by dashed lines. It is seen that the dashed line in diagrams c) and d) approximately coincides with the solid line for fragmentation mass $\sim 10^5 M_{\odot}$ given by numerical results of the collapsing filamentary cloud. Thus, we conclude that the condition $f < f_c$ with equation (37) provides a useful criterion for the formation of very massive fragments. Assuming equation (30) in the case with $J_{21}=1$, the criterion $f_c$ has about factor 3 of error.

%%%%%%%%%%%%%%%%%%%%%%%%%%%%%%%%%%%%%%%%%%%%%%%%%%%%%%%%
%\section{Rarefaction model}
%\section{Modified one-zone model with run-away collapse}
%\section{Modified one-zone model with run-away collapse}
\section{Effect of rarefaction wave}
%%%%%%%%%%%%%%%%%%%%%%%%%%%%%%%%%%%%%%%%%%%%%%%%%%%%%%%%
So far, we have assumed a uniform filamentary cloud where density of cloud is constant. However, the fragment mass predicted by the uniform model of the filamentary cloud tends to be lower than the result with more realistic treatment such as one-dimensional calculation (Uehara $\&$ Inutsuka 2000). Indeed, the fragment mass in \S 3 for the cases without the external radiation is different from that of Nakamura $\&$ Umemura (2002). During collapse, it is expected for the density profile to become core-envelope structure with uniform core and rarefied envelope. One physical reason to form such a density profile is the property of self-gravity. Central dense region collapses in a shorter time than outer less dense region, and density contrast increases. This is important for a cold collapsing cloud without pressure. While with the effect of pressure, run-away collapse is enhanced even if the initial cloud is uniform. Even without density perturbations, pressure gradient force erodes the filamentary cloud from the surface during collapse. Rarefaction wave propagates from the outer boundary to the center. Region outside the rarefaction wave front is delayed to collapse by the effect of outward pressure gradient force. This becomes important in a cloud with non-negligible pressure. In this section, to capture the effect of run-away collapse partly, the effect of run-away collapse which is induced by rarefaction wave is taken into account in the one-zone model of the collapsing filamentary cloud. Similar approach is adopted in a rotating isothermal cloud and is shown to be effective (Tsuribe $\&$ Inutsuka 1999). To include full characteristics of run-away collapse, the one-dimensional hydrodynamical calculations are required. Results of series of the one-dimensional calculations will be reported elsewhere.

%=========================================%
%\subsection{Propagating rarefaction wave}
\subsection{Modification to the model}
%=========================================%
Consider a collapsing filamentary cloud with uniform initial density and pressure. As the cloud collapses, radius of the cloud decreases. In addition to this, the rarefaction wave propagates inward from the outer boundary according to
\begin{eqnarray}
\frac{d\tilde{l}}{dt} = - 2 \pi \tilde{r} \rho c_s,
\end{eqnarray}
where $\tilde{r}$ is the position of the rarefaction front, $\rho$ and $\tilde{l}$ is the line mass and density inside $\tilde{r}$. Combining with the solution of density, velocity for homologous collapse, and $\tilde{l} = \pi \tilde{r}^2 \rho$, we can calculate the evolution of $\tilde{r}$ and $\tilde{l}$. Using $\tilde{r}$ and $\tilde{l}$, we define and solve the modified virial equation instead of equation (5) as
\begin{eqnarray}
\frac{d\tilde{v}}{dt}=-\frac{2G}{\tilde{r}} \{ \tilde{l}-l_c(T) \},
\end{eqnarray}
where $\tilde{v}$ is the infall velocity at the rarefaction wave front. Hereafter we denote this model as ``rarefied filament model". Different from the uniform model in previous sections, in the rarefied filament model $\tilde{l}$ decreases as the cloud collapses. Thus, in this model the right hand side of equation (39) becomes positive at lower density than the uniform model. Similarly, fragmentation occurs at lower density in the rarefied filament model. These differences are originated from the property of run-away collapse. The same condition as in \S 2.3 for fragmentation is assumed. Mass of the fragment is calculated using $M_{\mathrm{frag}} = 2 \pi \tilde{r} \tilde{l}$ instead of equation (18).

%===================%
\subsection{Results}
%===================%
%------------------------------------------------------------------%
%\subsubsection{Low-density filamentary clouds with large line mass}
%------------------------------------------------------------------%
In figure 13, the result for the rarefied filament model is shown for the case with large line mass and low initial density, $(f,n_0,J_{21})=(3,10\mathrm{cm^{-3}},0)$. Compared with the uniform model in figure 1, the cloud fragments at lower density as expected. Density at fragmentation is $5.0 \times 10^{-12}$ times that in the uniform model and effective radius of the filamentary cloud at the moment of fragmentation is larger by $5.1 \times 10^5$. On the other hand, line mass is $\tilde{l} = 2.1 \times 10^{-2} l$ at the moment of fragmentation. As a result of combination of these effects, the fragmentation mass ($127M_\odot$) is $1.1 \times 10^3$ time larger than in the uniform model. It should be noted that even with the rarefied filament model complete property of run-away collapse can not be captured for the filamentary cloud with initial density profile with central concentration. In this sense, the result of fragment mass still differs by about factor 4 from the previous result of the one-dimensional calculation (e.g., Nakamura \& Umemura 2002)\footnote{Note also that fragmentation timescale of Nakamura $\&$ Umemura (2002) is longer by factor $2.5$ than ours.}.

%-----------------------------------------------------------------%
%\subsubsection{High-density filamentary clouds with small line mass}
%-----------------------------------------------------------------%
In figure 14, the result is shown for the case with small line mass and high initial density, $(f,n_0,J_{21})=(1.25,10^6\mathrm{cm^{-3}},0)$. Compared with the uniform model in figure 2, in the rarefied filament model, fragmentation density is $3.2\times 10^{-2}$ of that in the uniform model. Effective radius of the filamentary cloud at the moment of fragmentation is $5.6$ times larger, but line mass at that moment is $0.25 l$. As a result, the fragment mass ($64M_\odot$) is $1.4$ times larger than in the uniform model.

Like figure 3, figure 15 shows the fragment mass for various $n_0$ and $f$ in the case without the external radiation. All lines are similar to each other and can be approximated as $M_{\mathrm{frag}} \sim 30000 n_0^{-0.3} f^{-4}$ with an error at most factor 2 at $f=3$. Although the fragment mass is determined mainly by $f$, dependence on $n_0$ is stronger than the case in figure 3. This tendency agrees with the result of Nakamura $\&$ Umemura (2002).

%-----------------------------------------------------------------%
%\subsubsection{Low-density filamentary clouds with small line mass and strong radiation}
%-----------------------------------------------------------------%
In figure 16, the result is shown for the case with small line mass, high initial density, and strong external radiation, $(f,n_0,J_{21})=(1.25,10\mathrm{cm^{-3}},10)$. Compared with the uniform model in figure 5, in the rarefied filament model, fragmentation density is $0.65$ of that in the uniform model. Effective radius of the filamentary cloud at the moment of fragmentation is $0.8$ of the uniform filamentary cloud and the line mass at that moment is $0.29 l$. As a result, the fragment mass is $0.23$ of that in the uniform model. Thus, the fragment mass is smaller ($\sim 2.6 \times 10^4 M_\odot$) due to smaller line mass, but still larger than without the external radiation. In all of above results, thermal evolution is qualitatively similar to that of the uniform model.

%=========================%
\subsection{Fragment mass}
%=========================%
In figure 17, contours maps of the fragment mass in $n_0 - f$ plane are shown for the cases with $J_{21}=0$, $1$, $6.5$, and $10$. Solid lines in each diagram represent constant fragment mass. Comparing diagram b), c), and d) with diagram a), it is seen that the fragment mass for the cases with low initial density ($n_0<10^{2-2.5} \mathrm{cm^{-3}}$) is strongly affected by the external radiation. In the cases with these low $n_0$, fragment mass is mainly determined by $n_0$ instead of $f$ and massive fragments form in the cases with $J_{21} \ge 1$. In the cases with high initial density ($n_0>10^3 \mathrm{cm^{-3}}$), the fragment mass is approximately independent of the external radiation. Since density is high enough for the filamentary cloud to shield itself from the dissociation photon from the early stage of collapse, role of photodissociation is less important.

We comment on the effects of run-away collapse. By comparing figure 17 with figure 10, these effects is clearly noticed. The most remarkable difference is that strong dependence on $f$ in figure 10 becomes weaker in figure 17. Furthermore, sub-solar fragments seen in figure 10 are not found in figure 17. Thus, we suspect that too small mass of fragments in figure 10 are the result of too idealized modeling with the uniform filamentary cloud in previous sections. In both of figure 17 and figure 10 massive fragments are seen in the cases with low initial density ($n_0<10^2\mathrm{cm^{-3}}$) and $J_{21}$ ($>1$). Formation of massive fragment with the external radiation can be regarded as a robust result.

%%%%%%%%%%%%%%%%%%%%%%%%%%%%%%%%%%%
%\section{Conclusion and discussion}
\section{Summary and discussion}
%%%%%%%%%%%%%%%%%%%%%%%%%%%%%%%%%%%
\subsection{Summary}
In this paper, we investigated collapse and fragmentation of primordial filamentary clouds under the external radiation with one-zone models. We numerically calculated the thermal and dynamical evolution of the filamentary clouds and estimated the mass of fragments for a variety of parameters such as $n_0$, $f$, and $J_{21}$. According to the uniform model it is found that with initial $\mathrm{H_2}$ fraction $f_{\mathrm{H_2}}=10^{-4}$, low initial density ($n_0 \le 10^2 \mathrm{cm^{-3}}$), and moderate line mass ($f \le 2$) the filamentary cloud loses its cooling ability as a result of photodissociation of $\mathrm{H_2}$ by the external radiation whose mean intensity is $J_{21} \ge 6.5$. In such a case, gravitational collapse proceeds adiabatically, and the filamentary clouds fragment into more massive fragments ($\sim 10^{4-5} M_\odot$) than the case without the external radiation ($\sim 1-50 M_\odot$). In the cases with lower intensity of the external radiation, the filamentary cloud collapses without fragmentation to density which is high enough for $\mathrm{H_2}$ to form as a result of self-shielding. In this case, mass of fragments is expected to be similar to the case without the external radiation. If the initial density is high ($n_0 > 10^2 \mathrm{cm^{-3}}$), the filamentary clouds with moderate line mass shields themselves from the dissociation photons. However, in such a high initial density case, adiabatic heating dominates cooling. As a result, they fragment into more massive fragments ($\sim 100 M_\odot$) than the low initial density cases with effective $\mathrm{H_2}$ cooling. Summarizing the results of numerical calculations, figure 10 clearly shows that the effect of the external dissociation radiation increases the fragment mass in low initial density cases (see $\S 3.3$). In $\S 4$, we derived an analytic criterion for the formation of very massive fragments via photodissociation. It is found that massive fragment is expected if the cooling time with equilibrium $\mathrm{H_2}$ fraction is longer than the free-fall time at the end of hypothetical adiabatic collapse.

In order to modify an unrealistic property of the uniform model where collapse is not suppressed until too high density (e.g., $\sim 10^{15} \mathrm{cm^{-3}}$ in figure 1), we developed a modified version of a simple one-zone model which can partly capture the effect of run-away collapse by focusing on the central dense spindle inside the rarefaction wave which comes from the outer boundary. According to this rarefied filament model, fragmentation is expected to occur before the rarefaction wave front arrives at the center. As a result, fragmentation density is smaller and fragment mass is usually larger than in the uniform model. With this new one-zone model, we can calculate easily the evolution of a filamentary cloud with run-away collapse without assuming free-fall collapse. Different from the uniform model, dependence of fragment mass on $f$ becomes weaker, and fragment mass itself becomes larger in the most cases in the $n_0 - f$ plane. This dependence and the value of fragment mass are similar to the result of the one-dimensional model (e.g., figure 5a of Nakamura $\&$ Umemura 2002). In the uniform model and the rarefied filament model, very massive fragments ($\ge 10^4 M_\odot$) form from the filamentary clouds with low initial density ($n_0 \le 10^2 \mathrm{cm^{-3}}$) and the external radiation. This formation of very massive fragments is the robust result.

\subsection{Discussion}
%% Dissociation property : Filament vs Sphere %%
We compare the difference of the effect of photodissociation between the spherical cloud (Omukai 2001) and the filamentary cloud with radius $R$. For the uniform spherical cloud, we have $N_{\mathrm{H_2}}=n_{\mathrm{H_2}} R \propto n_{\mathrm{H_2}} n^{-1/3} \propto n^{2/3}$ and for the uniform filamentary cloud we have $N_{\mathrm{H_2}}=(\pi/2) n_{\mathrm{H_2}} R \propto n_{\mathrm{H_2}} n^{-1/2} \propto n^{1/2}$ (see $\S 2.2$). Since the photodissociation reaction rate $k_{\mathrm{2step}}$ is proportional to $N_{\mathrm{H_2}}^{-3/4}$, we have
\begin{eqnarray}
k_{\mathrm{2step}} \propto \left\{ \begin{array}{ll}
n^{-1/2} & \mathrm{sphere} \\
n^{-3/8} & \mathrm{filament}. \\
\end{array} \right.
\end{eqnarray}
%Since $k_{\mathrm{2step}}$ of the collapsing uniform sphere decreases faster than that of the collapsing uniform filamentary cloud, mass of the uniform sphere correspond to $f_c$ in equation (38) is a little smaller than $f_c$. However, 
Difference in power index in both cases is small and the evolution of temperature would be similar to each other.

%%% Possible evolution of fragments and their effect %%%
We comment on the further evolution of very massive fragments ($\sim 10^{4-5}M_\odot$) which form as a result of fragmentation of the filamentary cloud under the external radiation. Since each fragment is expected to be nearly spherically symmetric, evolution of spherical cloud under the external dissociation radiation will be useful to discuss further evolution of each fragment. Susa (2007) investigated collapse of spherical cloud ($\sim 10^5 M_\odot$) under the UV radiation. When the distance between the cloud and a single light source ($120 M_\odot$ star) is longer than $100 \mathrm{pc}$, and the light source is turned on when the density of the cloud is $10^2 \mathrm{cm^{-3}}$, the author showed that the cloud collapses to $10^{6-7} \mathrm{cm^{-3}}$. Although our model adopts the uniform external radiation, similar evolution may be possible if the radiation field can be regarded as the effect from many light sources. In this case, mean intensity depends on the mean distance between the filamentary cloud and light sources. In our model with $J_{21}=6.5$, the distance between the filamentary cloud in a halo and surrounding star-forming halos (light sources) whose luminosity is $\sim 10^{25} \mathrm{erg\ s^{-1}}$ is expected to be longer than $20\mathrm{kpc}$ according to Dijkstra et al. (2008). Thus, mean intensity at the surface of the filamentary cloud is expected to be weaker than that of the situation of Susa (2007), and the external radiation is not expected to photodissociate $\mathrm{H_2}$ once it collapses to high density ($\sim 10^{3-4}\mathrm{cm^{-3}}$). Such a fragment may continue to collapse up to $10^{6-7} \mathrm{cm^{-3}}$ and shields itself from the external radiation. After that, although $\mathrm{H_2}$ will form in the collapsing clump and it cools the clump, other effect than cooling physics such as rotation and disk formation/accretion may be important. Since the possibility of further fragmentation will depends on these processes, the final outcome of each clump is out scope of the present paper. Assuming that each clump does not fragment further after it fragment at around the loitering point, Omukai $\&$ Yoshii (2003) discussed the initial mass function.

In the realistic situation, the effect of photodissociation is expected to be dominated by the single nearest point source (Susa et al. 2009; Hasegawa et al. 2009a). In this paper, however, we assume isotropic and steady external radiation field for simplicity. If this source of the external radiation is regarded as the group of sources (Dijkstra et al. 2008), new stars must be formed continuously around the filamentary cloud. Even in such a case, intensity and spectrum may be more complex and evolve with time. Furthermore, the effect from ionization photon may not be neglected. These issues may be important but are out scope of this paper.

%% Problem left for the future papers 1 : 1D or higher in hydro %%
Although the rarefied filament model is developed in this paper, it is still questionable whether or not all of the filamentary cloud fragment before the moment when rarefaction wave reaches the center. To clarify this point, more accurate calculation at least one-dimensional hydrodynamical calculation is required.

\bigskip

We thank Fumio Takahara for fruitful discussion and continuous encouragement, Kazuyuki Omukai for showing detailed technical treatment used in Omukai (2001), Shu-ichiro Inutsuka for discussion about the rarefied filament model. We also acknowledge the referee for improving the manuscript.

\appendix
\section{The opacity of species including deuterium}
According to Omukai (2001), opacity is determined by cross section, partition function, and reduced mass. Cross section is the same one that we consider in photoreactions. Reduced mass is estimated easily. According to Bron et al. (1973), the ratio of the partition function of $\mathrm{HD}$ to that of $\mathrm{H_2}$ is estimated. About $\mathrm{HD^+}$, the partition function $Z$ is
\begin{eqnarray}
Z=Z_{\mathrm{trans}}Z_{\mathrm{rot}}Z_{\mathrm{vib}}Z_{\mathrm{ele}},
\end{eqnarray}
where $Z_{\mathrm{trans}} \propto m^{3/2}$, $Z_{\mathrm{rot}} \propto I \propto m$, $ Z_{\mathrm{vib}} \propto m$, and $Z_{\mathrm{ele}} \propto m^0$. Hence $Z_{\mathrm{HD^+}}=(m_{\mathrm{HD^+}}/m_{\mathrm{H_2^+}})^{7/2} Z_{\mathrm{H_2^+}}$.

\section{The integral value of equation (14)}
The integrated value of equation (14) can be approximated by
\begin{eqnarray}
S(x)=\left\{ \begin{array}{ll}
4 \pi & x<0.01 \\
-4.10915x^3+12.579x^2-17.5787x+12.5392 & 0.01<x<1 \\
3.48330x^{-0.398517x-0.884854} & 1<x<5 \\
7.57551x^{-0.264805x-1.93051} & 5<x<10 \\
18.3238x^{-0.236295x-0.258468} & 10<x<20 \\
48.3927x^{-0.213846x-3.28825} & 20<x<30 \\
54.4202x^{-0.202620x-3.61369} & 30<x<40 \\
54.8962x^{-0.201651x-0.361636} & 40<x \\
\end{array} \right.
\end{eqnarray}
where $x \equiv k_\nu R$. Error is smaller than 10$\%$.

\section{Free-fall time of the uniform filamentary cloud}
Neglecting the effect of kinetic energy and pressure in equation (1) with
\begin{eqnarray}
I=\int_V \rho r^2 dV=\frac{1}{2}l R^2 ,
\end{eqnarray}
we have
\begin{eqnarray}
\frac{1}{2} \frac{d^2 \ }{dt^2} \biggl( \frac{1}{2}l R^2 \biggr)&=&-Gl^2 \\
\frac{d^2 \ }{dt^2} R^2 &=&-4Gl,
\end{eqnarray}
where $l$ is constant. From equation (A5), $R^2$ is the quadratic function of $t$. Hence we can express $R^2=at^2+bt+c$ ($a$, $b$, and $c$ are constant). When $t$ equals zero, $R$ is the initial radius and $c=R_0^2$. We assume that the initial velocity is zero and $b=0$. According to equation (A5), $a=-2Gl$. Using the relation $R_0=\sqrt{l/\pi \rho}$, we have
\begin{eqnarray}
R=\sqrt{-2Gl t^2+l/\pi \rho}.
\end{eqnarray}
Hence the free-fall time for the filamentary cloud is
\begin{eqnarray}
t=\frac{1}{\sqrt{2\pi G \rho}} .
\end{eqnarray}

%%%
% See the manual for the detail.
%%%

\newpage

\begin{table}[htbp]
\scalebox{0.7}[0.7]{
\begin{tabular}{c|c|c|c|c}
\hline
Number & Name & Process & Cross Section ($\mathrm{cm^{-2}}$) & Reference \\ \hline \hline
a1 & $\mathrm{H}$ bound-free & $\mathrm{H}(n) + \gamma \rightarrow \mathrm{H^+} + e$ & $7.909\times10^{-18} n (\nu/\nu_n)^{-3}$;$h\nu_n=13.598 \mathrm{eV}/n^2$ & 1 \\
a2 & $\mathrm{He}$ bound-free & $\mathrm{He} + \gamma \rightarrow \mathrm{He^+} + e$ & $7.83\times10^{-18} [1.66(\nu/\nu_{\mathrm{T}})^{-2.05}-0.66(\nu/\nu_{\mathrm{T}})^{-3.05}]$;$h\nu_{\mathrm{T}}=24.586 \mathrm{eV}$ & 2 \\
a3 & $\mathrm{H^-}$ bound-free & $\mathrm{H^-} + \gamma \rightarrow \mathrm{H} + e$ & $10^{-18} \lambda^3 (1/\lambda-1/\lambda_0)^{3/2}f(\lambda)$, $\lambda_0=1.6419 \mu \mathrm{m}$, $f(\lambda)$ from equation(5) of reference & 3 \\
a4 & $\mathrm{H_2^+}$ bound-free & $\mathrm{H_2^+} + \gamma \rightarrow \mathrm{H} + \mathrm{H^+}$ & see table2 of reference & 4 \\
a5 & $\mathrm{H^-}$ free-free & $\mathrm{H} + e + \gamma \rightarrow \mathrm{H} + e$ & $k^{\mathrm{ff}}_{\lambda} (T) k_B T n_e$; $k^{\mathrm{ff}}_{\lambda}$ from equation(6) of reference & 3 \\
a6 & $\mathrm{H}$ free-free & $\mathrm{H^+} + e + \gamma \rightarrow \mathrm{H^+} + e$ & $3.692\times 10^8 g_{\mathrm{ff}}(\nu,T)\nu^{-3} T^{-1/2} n_e$;we take $g_{\mathrm{ff}}(\nu,T)=1$ & \\
a7 & $\mathrm{H_2}$-$\mathrm{H_2}$ CIA & $\mathrm{H_2}(v,J) + \mathrm{H_2} + \gamma \rightarrow \mathrm{H_2}(v^\prime,J^\prime) + \mathrm{H_2}$ & see figure1 of reference & 5 \\
a8 & $\mathrm{H_2}$-$\mathrm{He}$ CIA & $\mathrm{H_2}(v,J) + \mathrm{He} + \gamma \rightarrow \mathrm{H_2}(v^\prime,J^\prime) + \mathrm{He}$ & see figure2 of reference & 5 \\
a9 & $\mathrm{D}$ bound-free & $\mathrm{D} + \gamma \rightarrow \mathrm{D^+} + e$ & same as a1 & 6 \\
a10 & $\mathrm{HD^+}$ bound-free & $\mathrm{HD^+} + \gamma \rightarrow \mathrm{D} + \mathrm{H^+}$ & same as a4 & 6 \\
a11 & $\mathrm{HD^+}$ bound-free & $\mathrm{HD^+} + \gamma \rightarrow \mathrm{D^+} + \mathrm{H}$ & same as a4 & 6 \\
a12 & $\mathrm{D^-}$ bound-free & $\mathrm{D^-} + \gamma \rightarrow \mathrm{D} + e$ & same as a3 & 7 \\
s1 & $\mathrm{H}$ Rayleigh & $\mathrm{H} + \gamma \rightarrow \mathrm{H} + \gamma^\prime$ & $5.799 \times 10^{-29} \lambda^{-4}+1.422 \times 10^{-30} \lambda^{-6} + 2.784 \times 10^{-32} \lambda^{-8}$ & 8 \\
s2 & Tomson & $e + \gamma \rightarrow e + \gamma^\prime$ & $6.65 \times 10^{-25}$ & 1 \\
\hline
\end{tabular}
}
\caption{Continuum processes. The wave length $\lambda$ is in units of $10^{-6} \mathrm{cm}$. REFERENCES-(1)Rybicki $\&$ Lightman 1979; (2) Osterbrock 1989; (3) John 1988; (4) Stancil 1994; (5) Borysow, Jorgensen, $\&$ Zheng 1997; (6) Galli $\&$ Palla 1998; (7) Frolov 2004; (8) Kurucz 1970. a1-a8, s1, and s2 are considered in Omukai 2001.}
\end{table}

\begin{figure}
\begin{center}
\begin{tabular}{cc}
\resizebox{80mm}{!}{\includegraphics{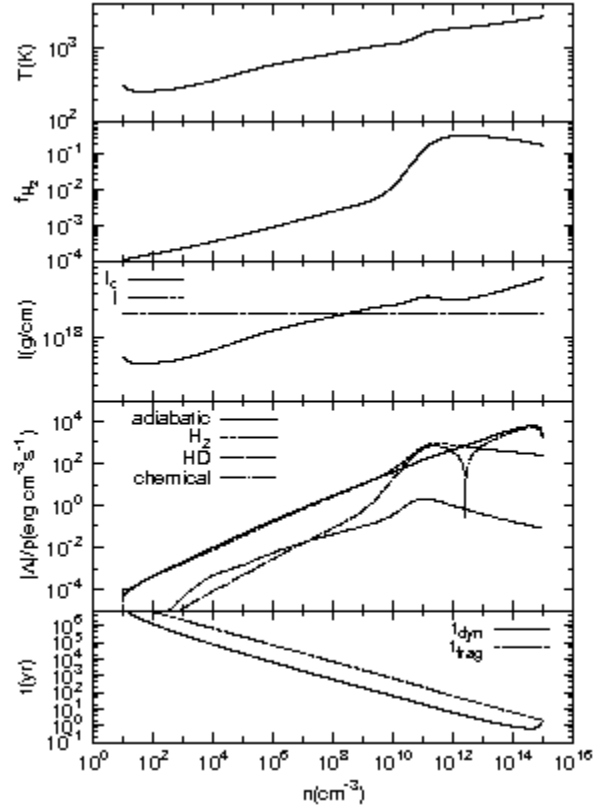}} \\
\end{tabular}
\caption{Evolution of the temperature (top), $f_{\mathrm{H_2}}$ (the second from top), $l$ and $l_c$ ($\S 2.1$:third), the heating and cooling rate (fourth), and $t_{\mathrm{dyn}}$ and $t_{\mathrm{frag}}$ ($\S 2.3$:bottom), respectively, as a function of the density for model with $(f,n_0,J_{21})=(3,10\mathrm{cm^{-3}},0)$, in which "adiabatic" denotes the adiabatic heating, "$\mathrm{H_2}$" does the $\mathrm{H_2}$ line cooling, "$\mathrm{HD}$" does the $\mathrm{HD}$ line cooling, and "chemical" does the chemical heating or cooling. We omit the continuum cooling because it is not effective.}
\end{center}
\end{figure}

\begin{figure}
\begin{center}
\begin{tabular}{cc}
\resizebox{80mm}{!}{\includegraphics{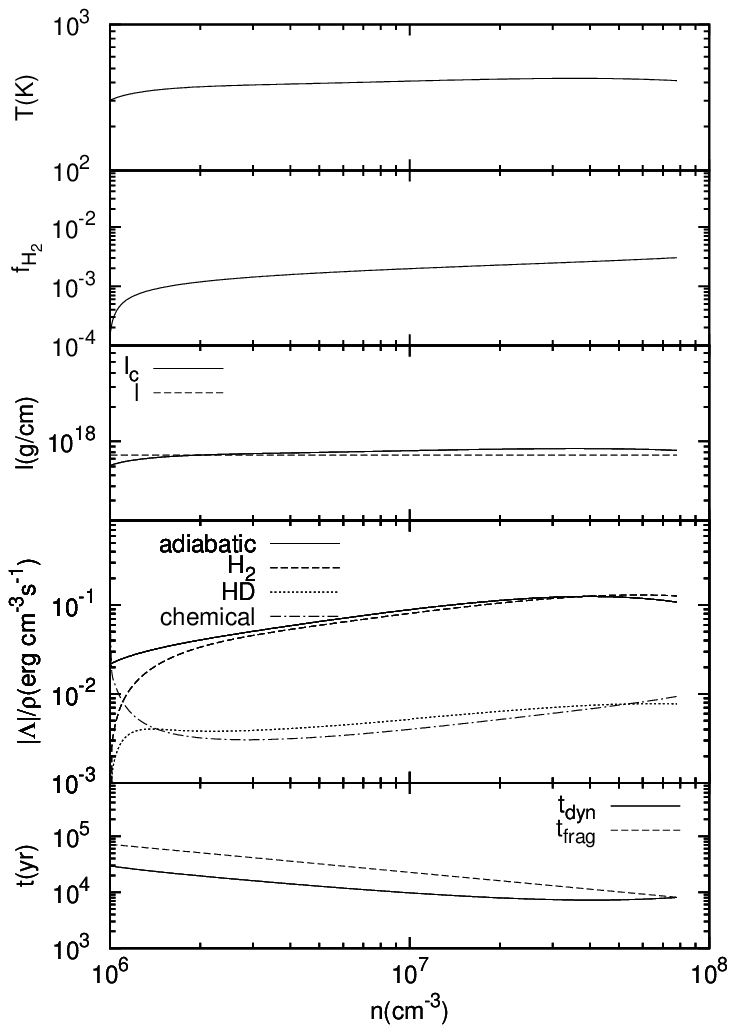}} \\
\end{tabular}
\caption{Same as figure1, but $(f,n_0,J_{21})=(1.25,10^6\mathrm{cm^{-3}},0)$.}
\end{center}
\end{figure}

\begin{figure}
\begin{center}
\begin{tabular}{cc}
\resizebox{80mm}{!}{\includegraphics{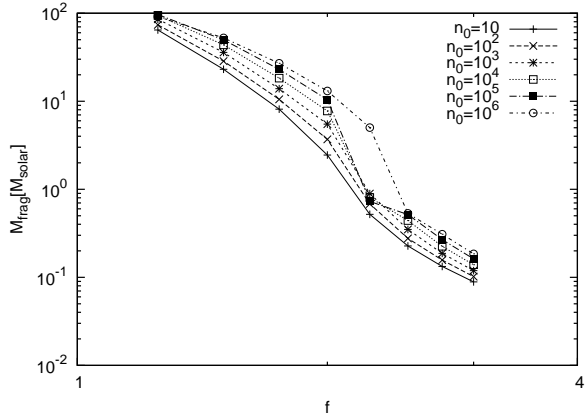}} \\
\end{tabular}
\caption{The fragment mass in the cases without the external radiation with various of $n_0$ and $f$. The hollowed region in the cases with $f=2.5$ and $n_0 \ge 10^5 \mathrm{cm^{-3}}$ in which $\mathrm{H_2}$ forms by three body reaction cools strongly. Since cooling helps collapse, the filamentary clouds collapse up to high density ($\sim 10^{13} \mathrm{cm^{-3}}$).}
\end{center}
\end{figure}

\begin{figure}
\begin{center}
\begin{tabular}{cc}
\resizebox{80mm}{!}{\includegraphics{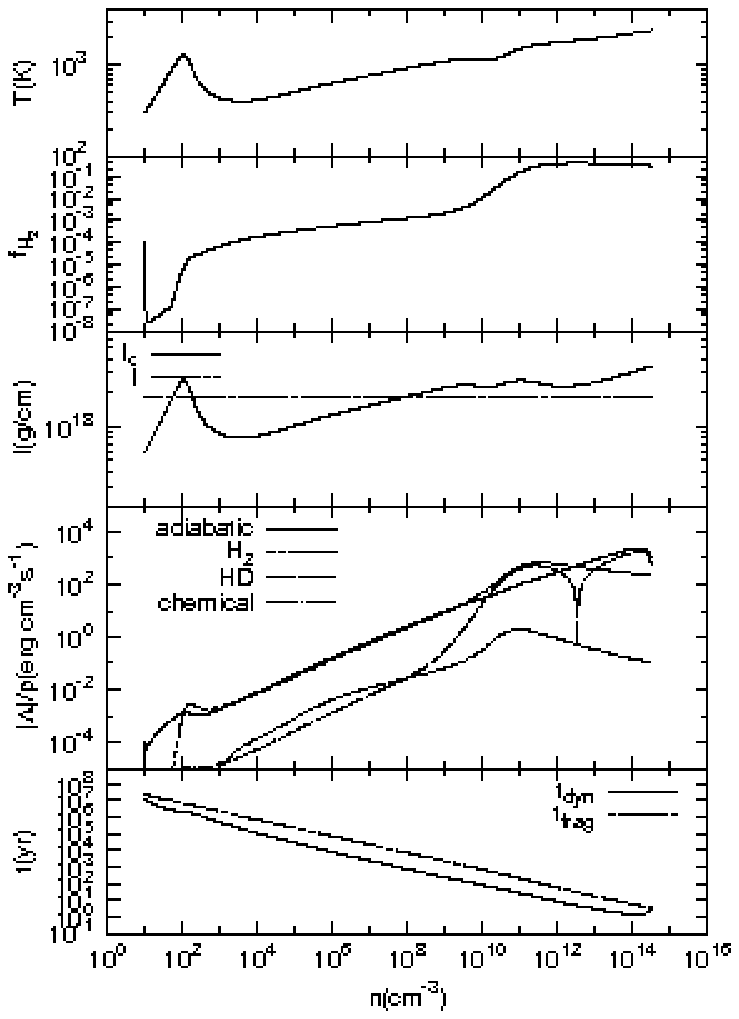}} \\
\end{tabular}
\caption{Same as figure1, but $(f,n_0,J_{21})=(3,10\mathrm{cm^{-3}},10)$.}
\end{center}
\end{figure}

\begin{figure}
\begin{center}
\begin{tabular}{cc}
\resizebox{80mm}{!}{\includegraphics{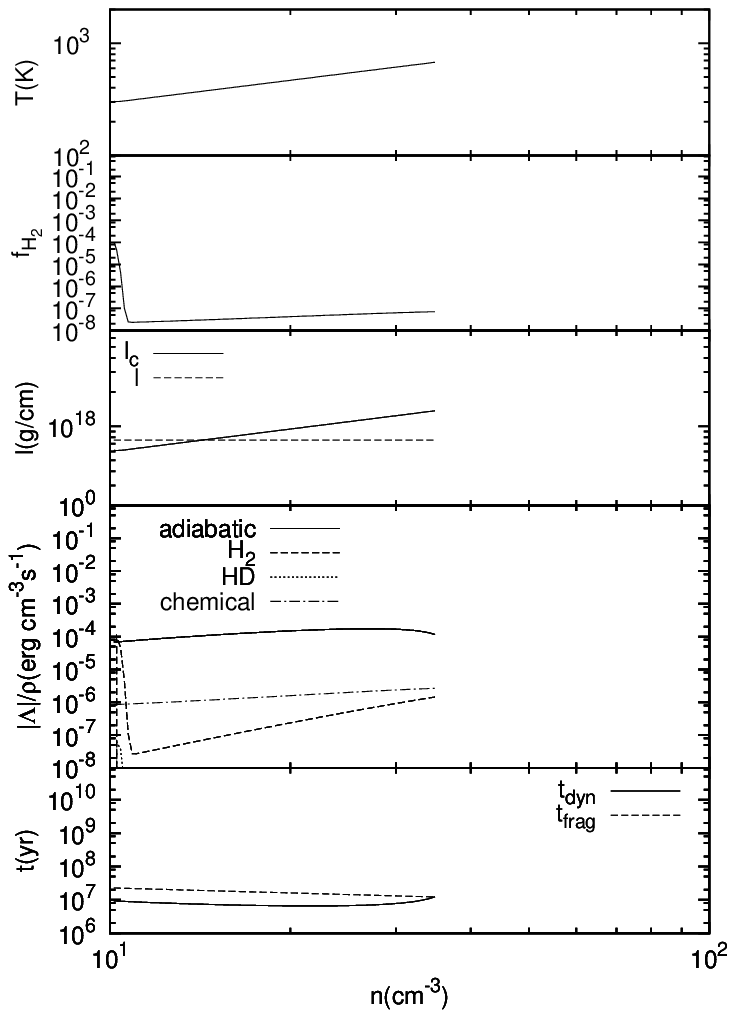}} \\
\end{tabular}
\caption{Same as figure1, but $(f,n_0,J_{21})=(1.25,10\mathrm{cm^{-3}},10)$.}
\end{center}
\end{figure}

\begin{figure}
\begin{center}
\begin{tabular}{cc}
\resizebox{80mm}{!}{\includegraphics{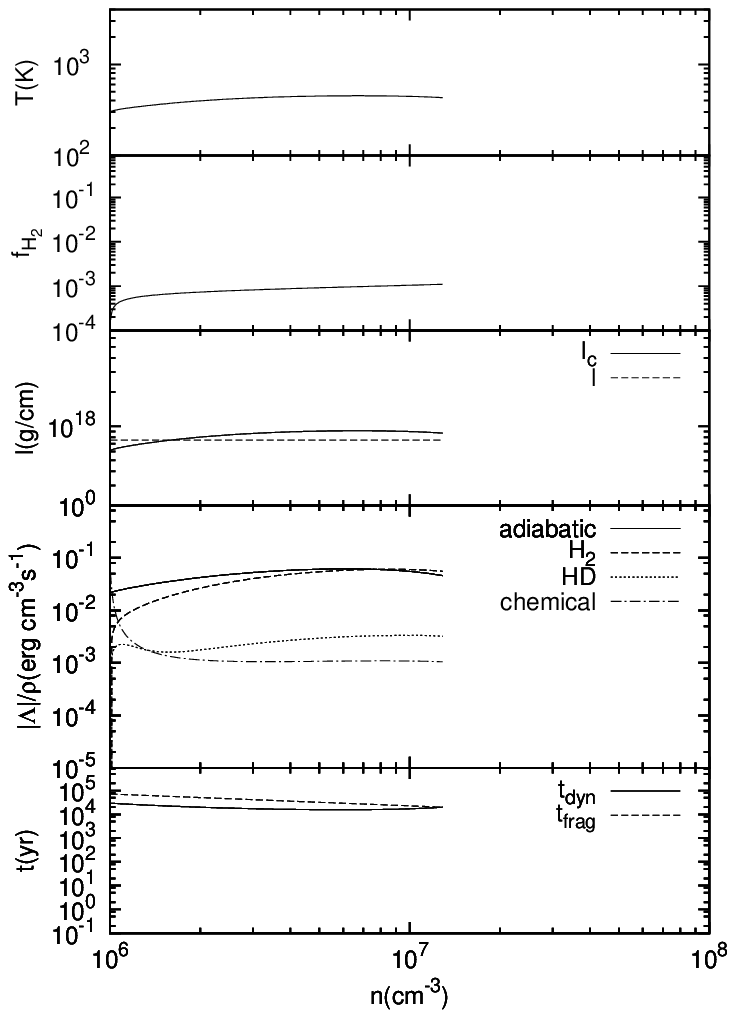}} \\
\end{tabular}
\caption{Same as figure1, but $(f,n_0,J_{21})=(1.25,10^6\mathrm{cm^{-3}},10)$.}
\end{center}
\end{figure}

\begin{figure}
\begin{center}
\begin{tabular}{cc}
\resizebox{80mm}{!}{\includegraphics{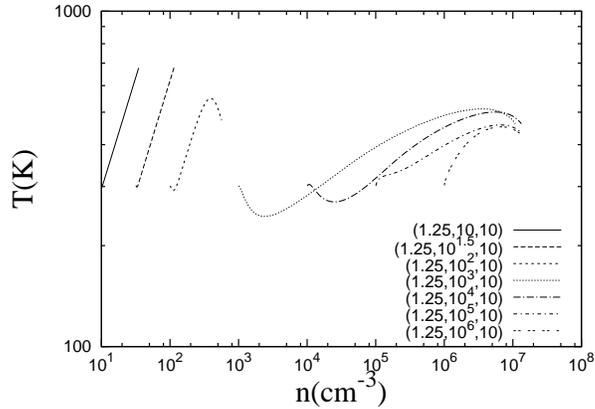}} \\
\end{tabular}
\caption{The evolution of the temperature for the cases with the variety of initial density, fixed $f$, and fixed $J_{21}$. The parentheses in the figure denotes the parameters $(f,n_0,J_{21})$. The right break in each line denotes $(T,n)$ at the fragmentation.}
\end{center}
\end{figure}

\begin{figure}
\begin{center}
\begin{tabular}{cc}
\resizebox{80mm}{!}{\includegraphics{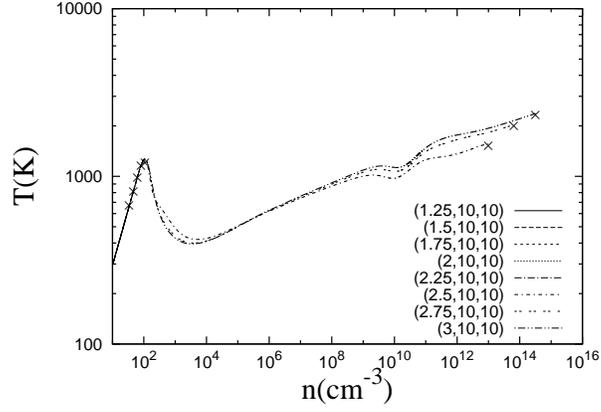}} \\
\end{tabular}
\caption{Same as figure7, but for the cases with the variety of line mass parameter, fixed $n_0$, and fixed $J_{21}$. Cross denotes $(T,n)$ at the fragmentation. It is clearly seen that fragmentation is divided into two groups with $n \sim 10^2 \mathrm{cm^{-3}}$ and $n \ge 10^{13} \mathrm{cm^{-3}}$.}
\end{center}
\end{figure}

\begin{figure}
\begin{center}
\begin{tabular}{cc}
\resizebox{80mm}{!}{\includegraphics{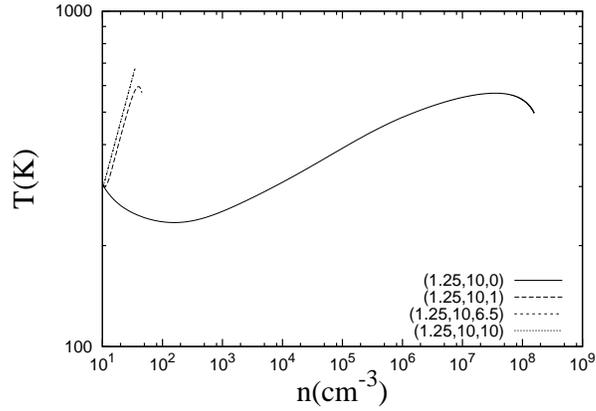}} \\
\end{tabular}
\caption{Same as figure7, but for the cases with the variety of intensity of the external radiation, fixed $n_0$, and fixed $f$. The lines for $J_{21}=6.5$ and $10$ overlap each other.}
\end{center}
\end{figure}

%\onecolumn

\begin{figure}
\begin{center}
\begin{tabular}{cc}
\resizebox{80mm}{!}{\includegraphics{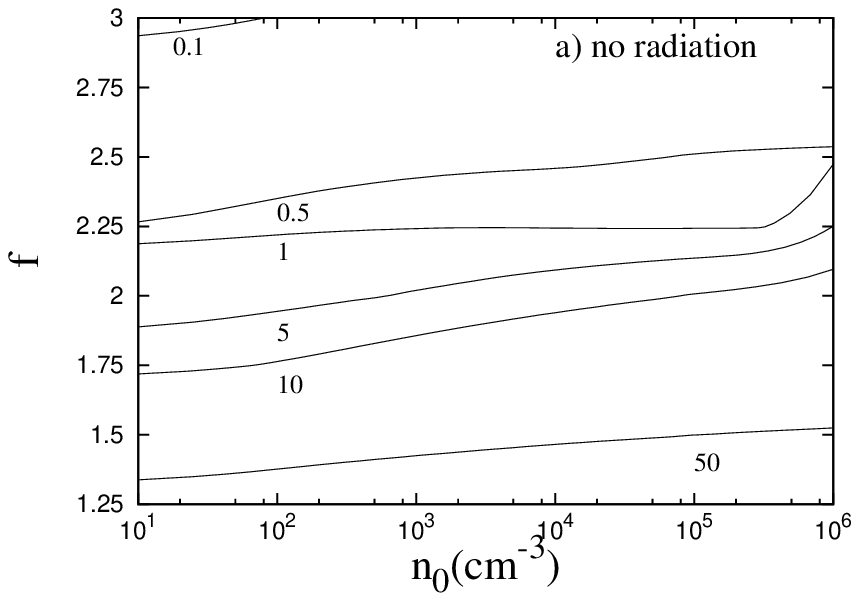}} &
\resizebox{80mm}{!}{\includegraphics{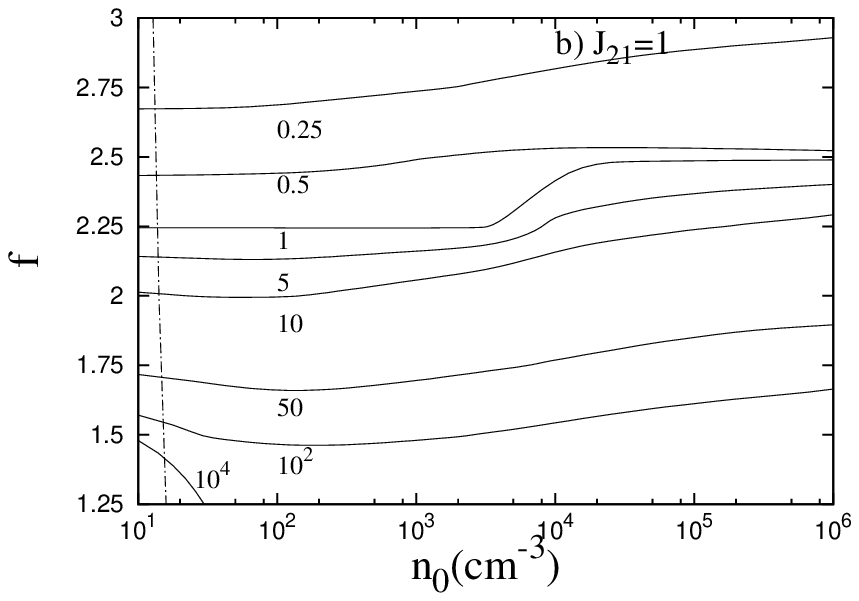}} \\
\resizebox{80mm}{!}{\includegraphics{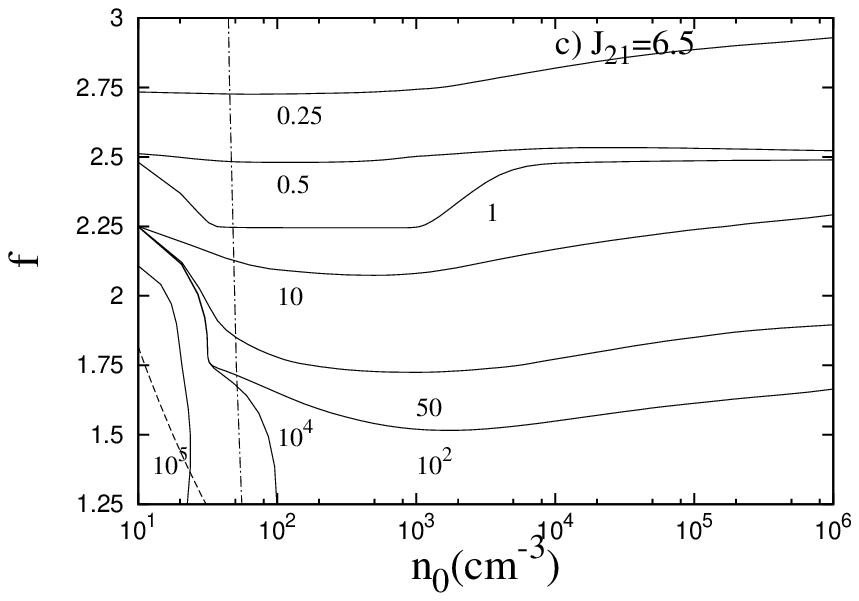}} &
\resizebox{80mm}{!}{\includegraphics{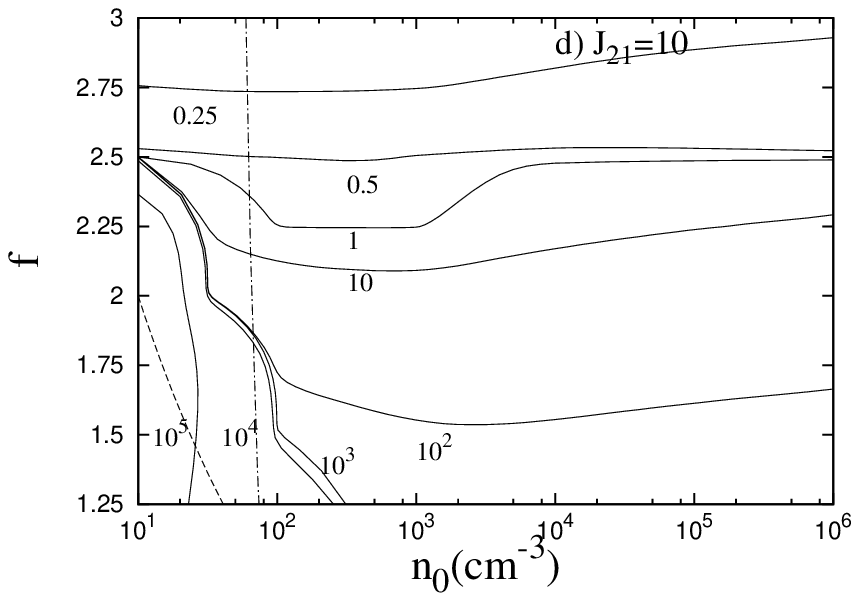}} \\
\end{tabular}
\caption{The contours map for the fragment mass in $n_0 -f$ plane for the case with a) $J_{21}=0$, b) $J_{21}=1$, c) $J_{21}=6.5$, and d) $J_{21}=10$. The number near each solid line is mass of fragment in units of $M_\odot$. The dashed line and the dot-dashed line denote equation (38) and equation (34), respectively. In the region on right of the dash-dotted line, $t_{\mathrm{cool}}>t_{\mathrm{ff}}$.}
\end{center}
\end{figure}

\begin{figure}
\begin{center}
\begin{tabular}{cc}
\resizebox{80mm}{!}{\includegraphics{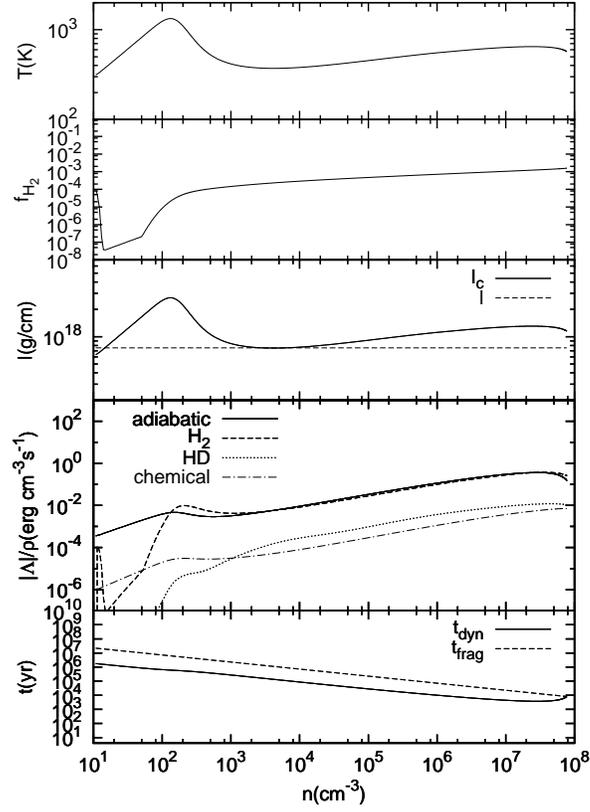}} \\
\end{tabular}
\caption{Same as figure1, but the initial velocity is five times of the sound speed and $(f,n_0,J_{21})=(1.25,10\mathrm{cm^{-3}},10)$.}
\end{center}
\end{figure}

\begin{figure}
\begin{center}
\begin{tabular}{cc}
\resizebox{80mm}{!}{\includegraphics{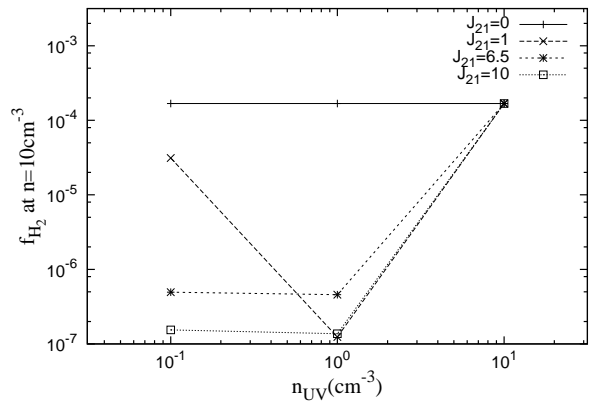}} \\
\end{tabular}
\caption{The fraction of $\mathrm{H_2}$ at $n=10\mathrm{cm^{-3}}$, as function of $n_{\mathrm{UV}}$ which is the density at which light sources turn on. Each line corresponds to various $J_{21}$.}
\end{center}
\end{figure}

\begin{figure}
\begin{center}
\begin{tabular}{cc}
\resizebox{80mm}{!}{\includegraphics{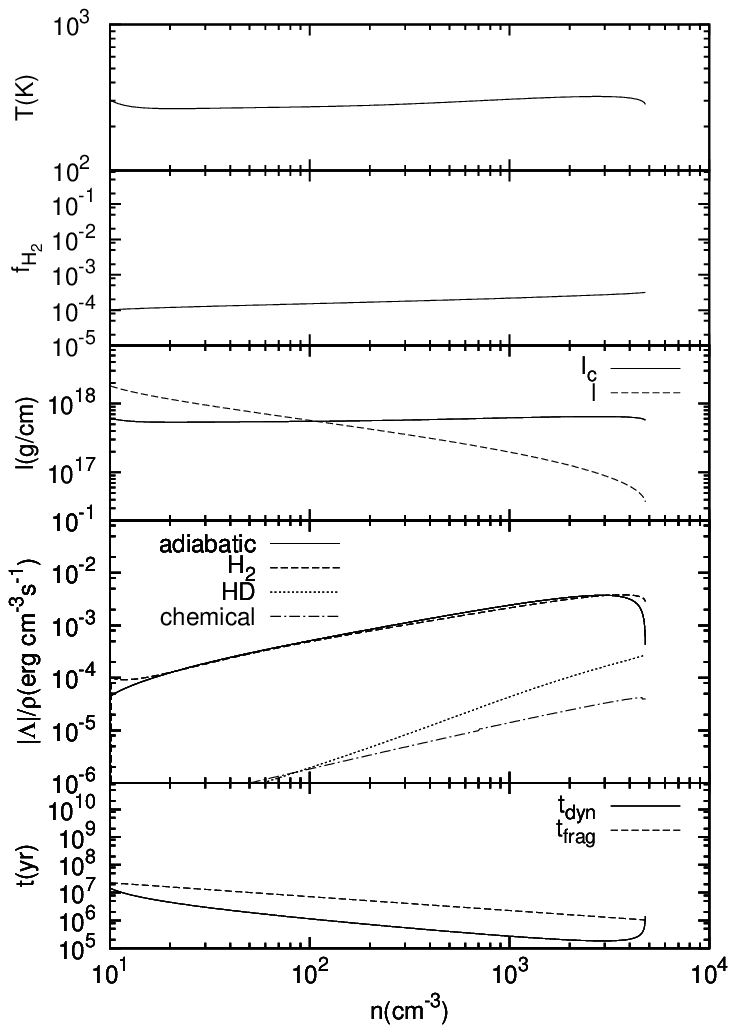}} \\
\end{tabular}
\caption{Same as figure1, but for the rarefied filament model, $(f,n_0,J_{21})=(3,10\mathrm{cm^{-3}},0)$.}
\end{center}
\end{figure}

\begin{figure}
\begin{center}
\begin{tabular}{cc}
\resizebox{80mm}{!}{\includegraphics{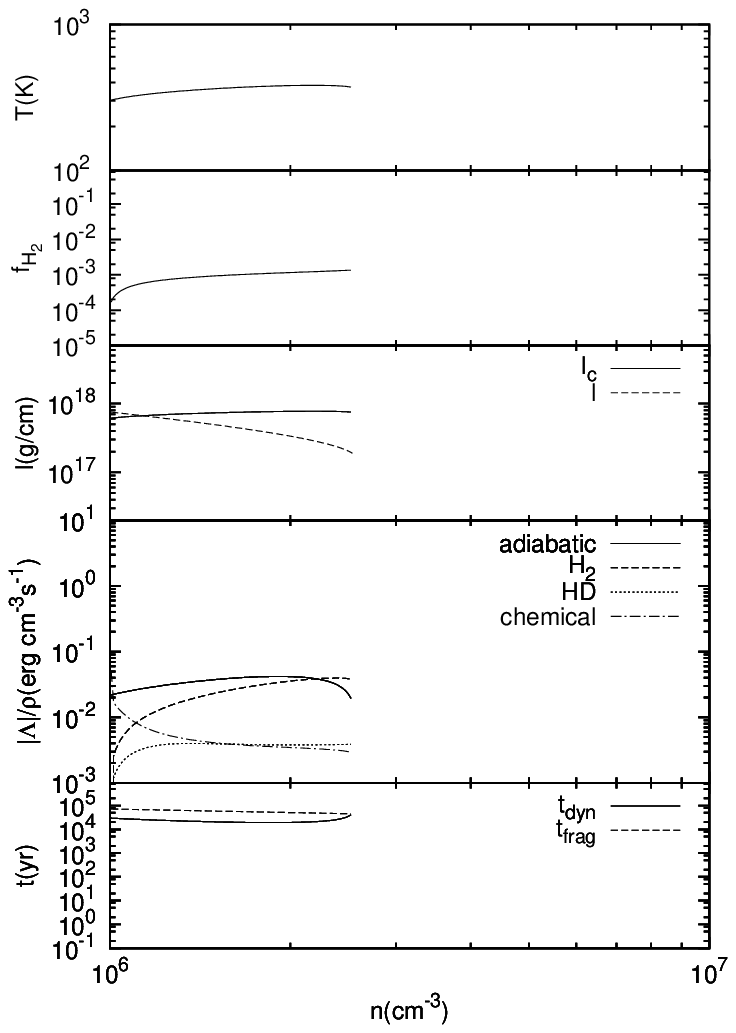}} \\
\end{tabular}
\caption{Same as figure2, but for the rarefied filament model, $(f,n_0,J_{21})=(1.25,10^6\mathrm{cm^{-3}},0)$.}
\end{center}
\end{figure}

\begin{figure}
\begin{center}
\begin{tabular}{cc}
\resizebox{80mm}{!}{\includegraphics{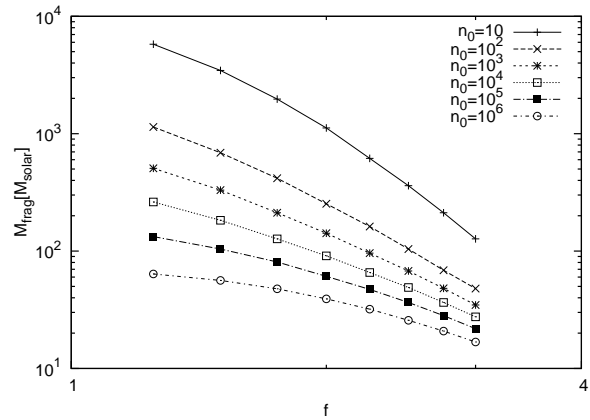}} \\
\end{tabular}
\caption{The fragment mass in the cases without the external radiation with various of $n_0$ and $f$ in rarefied filament model.}
\end{center}
\end{figure}

\begin{figure}
\begin{center}
\begin{tabular}{cc}
\resizebox{80mm}{!}{\includegraphics{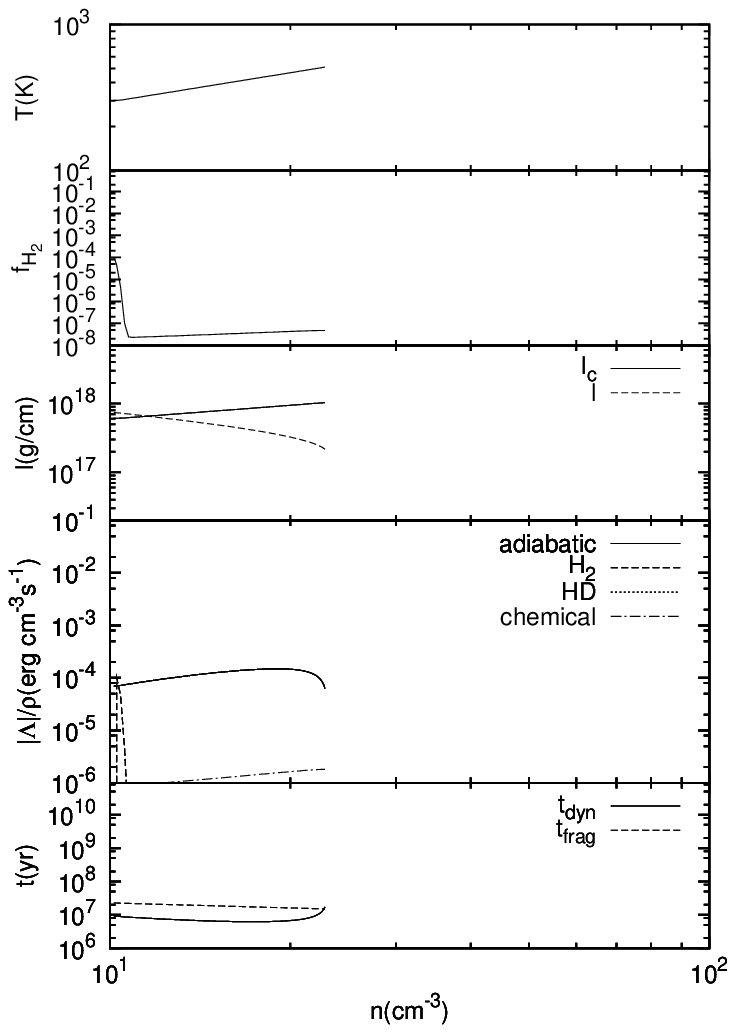}} \\
\end{tabular}
\caption{Same as figure5, but for the rarefied filament model, $(f,n_0,J_{21})=(1.25,10\mathrm{cm^{-3}},10)$.}
\end{center}
\end{figure}

\begin{figure}
\begin{center}
\begin{tabular}{cc}
\resizebox{80mm}{!}{\includegraphics{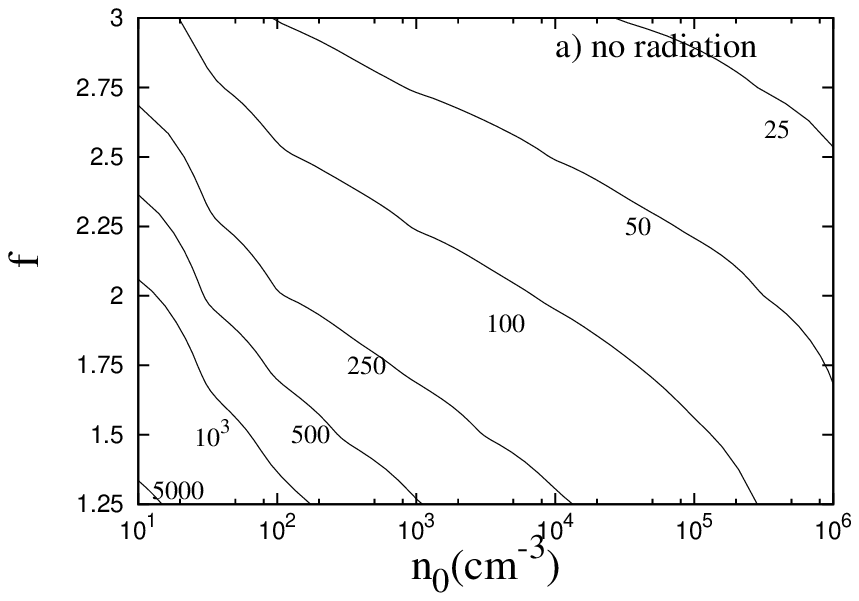}} &
\resizebox{80mm}{!}{\includegraphics{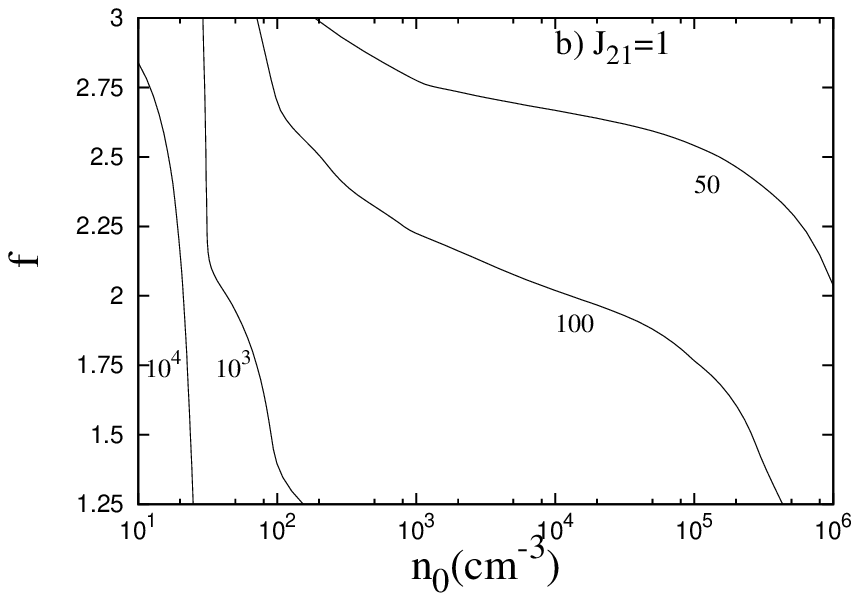}} \\
\resizebox{80mm}{!}{\includegraphics{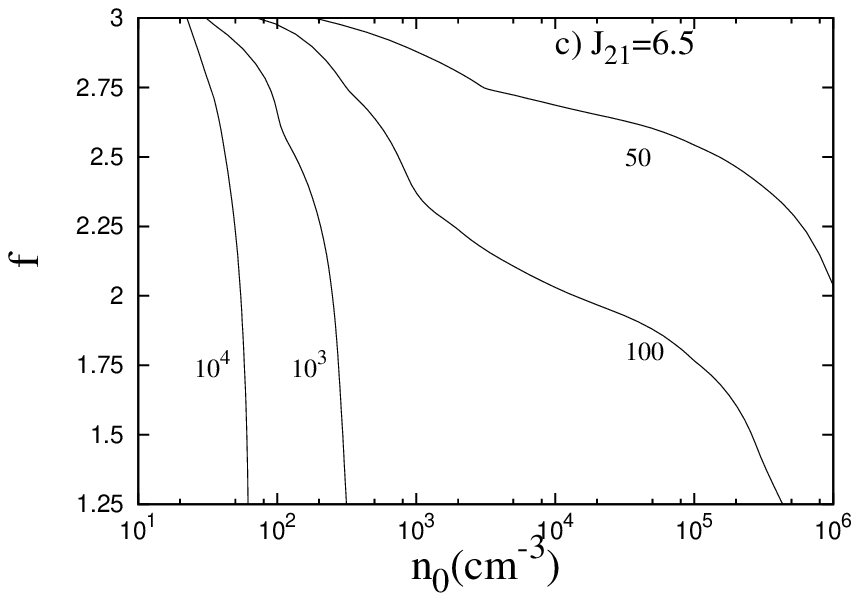}} &
\resizebox{80mm}{!}{\includegraphics{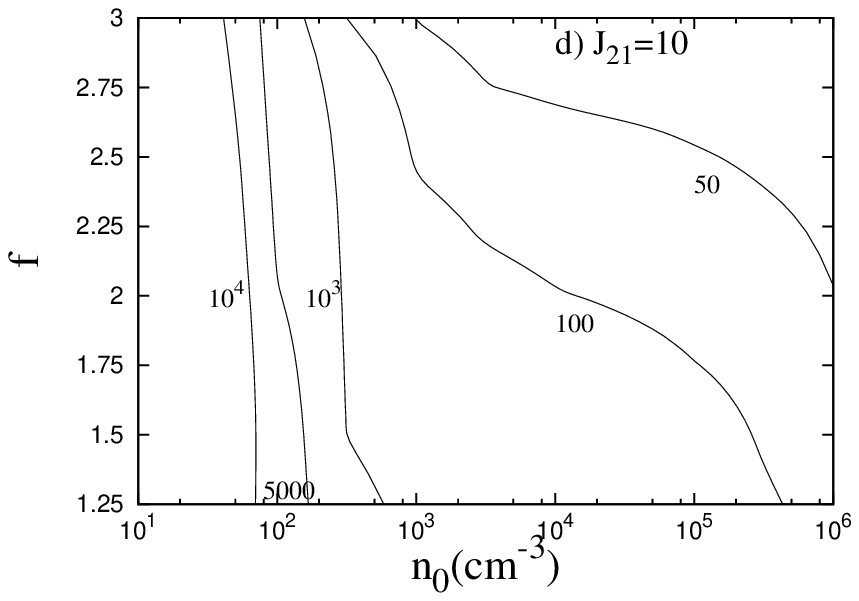}} \\
\end{tabular}
\caption{The contours map for the fragment mass for the case with a) $J_{21}=0$, b) $J_{21}=1$, c) $J_{21}=6.5$, and d) $J_{21}=10$. The number near each solid line is mass of fragment in units of $M_\odot$.}
\end{center}
\end{figure}

\end{document}